\documentclass[twoside,11pt]{article}

\usepackage{jmlr2e_arxiv}

\usepackage{chngcntr} 
\usepackage{rotating}
\usepackage{subcaption}
\usepackage{caption}
\usepackage{tikz} 
\usetikzlibrary{arrows,shapes}
\usepackage{bigints} 

\usepackage[utf8]{inputenc} 
\usepackage[T1]{fontenc}    
\usepackage{hyperref}       
\usepackage{url}            
\usepackage{booktabs}       
\usepackage{nicefrac}       
\usepackage{microtype}      
\usepackage{xspace}

\usepackage{xcolor} 
\usepackage[utf8]{inputenc}
\usepackage[english]{babel}
\usepackage{multicol}
\usepackage{fontawesome} 

\newtheorem{thm}{Theorem}
\newtheorem{lem}{Lemma}

\newcommand{\E}{\mathbb{E}}

\mathchardef\mhyphen="2D
\def\y{\mathbf{y}}

\def \E{\mathbb{E}}

\def \I {\mathbb{I}}

\def \dt {\mathrm{d}}

\newtheorem{assumption}{Assumption}
\newtheorem{rem}{Remark}

\ShortHeadings{Loss-Based Variational Bayes Prediction}{Frazier, Loaiza-Maya, Martin, and Koo}
\firstpageno{1}

\begin{document}

\title{Loss-Based Variational Bayes Prediction}

\author{\name David T. Frazier \email david.frazier@monash.edu \\
       \addr Department of Econometrics and Business Statistics\\ Monash
       University\\
       Melbourne, Australia, 3800
       \AND
       \name Rub\'en
       Loaiza-Maya \email ruben.loaizmaya@monash.edu \\
        \addr Department of Econometrics and Business Statistics\\ Monash
  University\\
  Melbourne, Australia, 3800
  	\AND   
\name Gael M. Martin \email gael.martin@monash.edu \\
      \addr Department of Econometrics and Business Statistics\\ Monash
University\\
Melbourne, Australia, 3800
\AND
\name Bonsoo Koo \email bonsoo.koo@monash.edu\\
      \addr Department of Econometrics and Business Statistics\\ Monash
University\\
Melbourne, Australia, 3800
}

\editor{}

\maketitle

\begin{abstract}
We propose a new approach to Bayesian prediction that caters for models with
a large number of parameters and is robust to model misspecification. Given
a class of high-dimensional (but parametric) predictive models, this new
approach constructs a posterior predictive using a variational approximation
to a generalized posterior that is directly focused on predictive accuracy.
The theoretical behavior of the new prediction approach is analyzed and a
form of optimality demonstrated. {Applications to both simulated and
	empirical data using high-dimensional Bayesian neural }network and{\
	autoregressive mixture models} demonstrate that the approach provides more
accurate results than various alternatives, including misspecified
likelihood-based predictions.
\end{abstract}

\begin{keywords}
Loss-based Bayesian forecasting;
variational inference; generalized (Gibbs) posteriors; proper scoring rules;
Bayesian neural networks; M4 forecasting competition
\end{keywords}

\section{Introduction}

The conventional paradigm for Bayesian prediction is underpinned by the
assumption that the true data generating process is either equivalent to the
predictive model adopted, or spanned by a finite set of models over which we
average. Of late however, recognition of the unrealistic nature of such an
assumption, allied with an increased interest in driving prediction by
problem-specific measures of accuracy, {or loss, }have led to alternative
approaches. Whilst antecedents of these new principles are found in the
`probably approximately correct' (PAC)-Bayes approach to prediction in the
machine learning literature (see \citealp{alquier2021user}, for a review),
it is in the statistics and econometrics literature that this `loss-based
prediction' has come to more formal maturity, including in terms of its
theoretical validation. This includes Bayesian work on weighted combinations
of predictions, such as, for example, \citet{Billio2013}, \citet{casarin2015}%
, \citet{Pett2016}, \citet{Bassetti2018}, \citet{BASTURK2019}, %
\citet{McAlinn2019}{ and \citet{McAlinn2020},} where weights are updated via
various predictive criteria, and the true model is not assumed to be one of
the constituent models - i.e. an $\mathcal{M}$-open state of the world
(Bernardo and Smith, 1994) is {implicitly }adopted. It also {includes a
	contribution} by \cite{loaiza2019focused}, in which both single models and
predictive mixtures are used to generate accurate Bayesian predictions in
the presence of model misspecification; with both theoretical and numerical
results highlighting the ability of the approach to out-perform conventional
likelihood-based Bayesian predictive methods.

However, as a general rule, the existing approaches discussed above do not scale well to complex models with high-dimensional parameter spaces. In contrast, the current paper contributes to the above literature by providing a new
approach for producing accurate loss-based predictions in high-dimensional
problems. {We begin by defining a class of }flexible{\ predictive models{, }}%
conditional on a set of unknown parameters, {that are a plausible mechanism
	for generating probabilistic predictions. A prior distribution is placed
	over }the parameters of this predictive {class, and the prior then updated }%
to a posterior{\textbf{\ }via a criterion function that captures a {%
		user-specified measure of predictive accuracy. That is, the conventional,
		and potentially misspecified likelihood-based update is eschewed, in favour
		of a function that is tailored to the predictive problem at hand; th}}e
ultimate{{\textbf{\ }goal being to produce accurate predictions according to
		the measure that }}matters, without requiring knowledge of the true data
generating mechanism.

In the spirit of the various generalized Bayesian \textit{inferential}
methods, in which likelihood functions are also replaced by alternative
updating mechanisms (\textit{inter alia}, \citealp{chernozhukov2003mcmc}, 
\citealp{Zhang2006a}{{,
		\citealp{Zhang2006b}, {\citealp{jiang2008}},
		{\citealp{bissiri2016general}}}}{{{{, }\citealp{giummole2017objective}}}},       
{{{\citealp{GVI2019}}}, }{\citealp{miller2019robust}, \citealp{Syring2019}}%
,\ and \citealp{pacchiardi2021generalized}), we adopt a coherent update
based on the exponential of a scaled sample loss; we refer to \cite%
{miller2021asymptotic} for a {thorough discussion of} the large sample
behavior of generalized Bayesian posteriors. As in \cite{loaiza2019focused}
the loss is, in turn, defined by a proper scoring rule \textbf{(}%
\citealp{gneiting2007probabilistic}; \citealp{gneiting2007strictly}\textbf{)}
that rewards a given form of predictive accuracy; for example, accurate
prediction of extreme values. Given the high-dimensional nature of the
resultant posterior, numerical treatment via `exact' Markov Chain Monte
Carlo (MCMC) is computationally challenging; hence we adopt an `approximate'
approach using variational principles. Since the posterior that results from
an exponentiated loss was first denoted {as }a `Gibbs posterior' by \cite%
{Zhang2006a},\textbf{\textbf{\ }}we refer to the variational approximation
of this posterior{\textbf{\ }}as the\textit{\ Gibbs variational posterior,}
and the predictive distribution that results from this posterior, via the
standard Bayesian calculus, as {the }\textit{Gibbs variational predictive }%
(hereafter, GVP). With a slight abuse of terminology, and when it is clear
from the context, we also use the abbreviation GVP to reference the method
of Gibbs variational prediction, or loss-based variational prediction \textit{per se.}

In an artificially simple `{toy' }example in which MCMC sampling of
the exact Gibbs posterior {is feasible, we illustrate that
	there are negligible }differences between the out-of-sample results yielded
by the GVP and those produced by the predictive based on MCMC sampling from
the {Gibbs} posterior. We establish this result under both correct
specification, in which the true data generating process matches the adopted
predictive model, and under misspecification of the predictive model. The {%
	correspondence} between the `approximate' and `exact' predictions in the
correct specification case mimics that documented in \cite%
{frazier2019approximate}, in which posterior approximations are produced by
approximate Bayesian computation (ABC) and for the log score update (only).\
In the misspecified case, `strictly coherent' predictions are produced (%
\citealp{martin2020optimal}), whereby a given GVP, constructed\ via the use
of a particular scoring rule, is shown to perform best out-of-sample
according to that same score when compared with a GVP constructed via some
alternative scoring rule; with the numerical values of the average
out-of-sample scores closely matching those produced by the exact Gibbs
predictive. That is, building a generalized posterior via a given scoring
rule yields superior predictive accuracy in that rule \textit{despite any
	inaccuracy} in the measurement of posterior uncertainty that is induced by
the variational approximation.

We then undertake more extensive Monte Carlo experiments to highlight the
power of the approach in genuinely high-dimensional problems, with
predictives based on: an autoregressive mixture model with 20 mixture
components, and a neural network model used for illustration. An empirical
analysis, in which GVP is used to produce accurate prediction intervals for
the 4227 daily time series used in the M4 forecasting competition,
illustrates the applicability of the method to reasonably large, and
realistic data sets.

While the {numerical toy example} {demonstrates} that little
predictive accuracy is lost when using the GVP, relative to the `exact'
predictive, we {also }rigorously compare the theoretical behavior {of 
}the GVP and the potentially infeasible {exact} predictive. Specifically, we demonstrate that in large samples the GVP delivers
	predictions that are just as accurate as those obtained from the exact Gibbs
	predictive when measured according to the score used to construct the Gibbs
	posterior. We do this by proving that the GVP `merges' (in the sense of \citealp{blackwell1962merging}) with the exact Gibbs predictive. This merging result relies on novel posterior concentration results that extend existing
	concentration results for generalized posteriors (\citealp{Alquier2016}, \citealp{alquier2020concentration}, and \citealp{yang2020alpha}) to
	temporally dependent, potentially heavy-tailed data.
	
The remainder of the paper is structured as follows. In Section \ref{setup}
the loss-based paradigm for Bayesian prediction is defined, with its links
to related segments of the literature (as flagged briefly above) detailed.
In Section \ref{GVP} we detail how to construct the GVP, and provide
theoretical verification of its accuracy. A numerical illustration of the
ability of the variational method to yield essentially equivalent predictive
accuracy to that produced via MCMC sampling is given in Section \ref{sec:toy}%
, using a low-dimensional example. The approach that we use in the
implementation of GVP, including the choice of variational family, is
briefly described, with further computational details of the stochastic
gradient ascent (SGA) method used to perform the optimization are included in a supplementary appendix to the paper. We then proceed with the numerical illustration of
the method in high-dimensional settings - using both artificially simulated
and empirical data - in Sections \ref{complex} and \ref{emp} respectively;
with the illustrations highlighting that, overall, the method `works' and
`works well'. We conclude in Section \ref{disc} with discussion of the
implications of our findings, including for future research directions. Proof of all theoretical results are given in an appendix, and all
computational details, and prior specifications are included in supplementary appendix to the paper.

\section{Setup and Loss-Based Bayesian Prediction\label{setup}}

\subsection{Preliminaries and Notation\label{prelim}}

Consider a stochastic process $\{Y_t:\Omega\rightarrow\mathcal{Y} , t\in%
\mathbb{N}\}$ defined on the complete probability space $(\Omega, \mathcal{F}%
,P_0)$. Let $\mathcal{F}_t:=\sigma(Y_1,\dots,Y_t)$ denote the natural
sigma-field, and let $P_0$ denote the infinite-dimensional distribution of
the sequence $Y_1,Y_2,\dots$. Let $y_{1:n}=(y_1,\dots,y_n)^{\prime }$ denote
a vector of realizations from the stochastic process.

Our goal is to use a particular collection of statistical models, adapted to 
$\mathcal{F}_{n}$, that describe the behavior of the observed data, to
construct accurate predictions for the random variable $Y_{n+1}$. The
parameters of the model are denoted by $\theta _{n}$, the parameter space by 
$\Theta _{n}\subseteq \mathbb{R}^{d_{n}}$, where the dimension $d_{n}$ could
grow as $n\rightarrow \infty $ and $\Theta _{1}\subseteq \Theta
_{2}\subseteq ...\subseteq \Theta _{n}$. For the notational simplicity, we
drop\ the dependence of\textbf{\ }$\theta _{n}$ and $\Theta _{n}$ on $n$ in
what follows. Let $\mathcal{P}^{(n)}$ be a generic class of one-step-ahead
predictive models for $Y_{n+1}$, conditioned on the information $\mathcal{F}%
_{n}$ available at time $n$, such that $\mathcal{P}^{(n)}:=\{P_{\theta
}^{(n)},\theta \in \Theta \}$.\footnote{{The treatment of scalar $Y_{t}$ and
		one-step-ahead prediction is for the purpose of illustration only, and all
		the methodology that follows can easily be extended to multivariate $Y_{t}$
		and multi-step-ahead prediction in the usual manner.}} When $P_{\theta
}^{(n)}(\cdot )$ admits a density with respect to the Lebesgue measure, we
denote it by $p_{\theta }^{(n)}(\cdot )\equiv p_{\theta }(\cdot |\mathcal{F}%
_{n})$. The parameter ${\theta }$ thus\textbf{\ }indexes values in the
predictive class, with ${\theta }$ taking values in the complete probability
space $(\Theta ,\mathcal{T},\Pi )$, and where $\Pi $ {measures} our beliefs
- {either prior or posterior - }about the unknown parameter ${\theta }$, and
when {they} {exist} we denote {the respective densities} by $\pi (\theta )$ {%
	and} $\pi (\theta |y_{1:n})$.

Denoting the likelihood function by $p_{\theta }(y_{1:n})$, the conventional
approach to Bayesian prediction updates prior beliefs about ${\theta }$ via
Bayes rule, to form the Bayesian posterior density,%
\begin{equation}
\pi (\theta |y_{1:n})=\frac{p_{\theta }\left( y_{1:n}\right) \pi (\theta )}{%
	\int_{\Theta }p_{\theta }\left( y_{1:n}\right) \pi (\theta )d\theta },
\label{exact_post}
\end{equation}%
in which we follow convention and abuse notation by writing this quantity as
a density even though, strictly speaking, the density may not exist. The
one-step-ahead predictive distribution is then constructed as 
\begin{equation}
P_{\Pi }^{(n)}:=\int_{\Theta }P_{\theta }^{(n)}\pi (\theta |y_{1:n})\dt%
\theta .  \label{conv_pred}
\end{equation}%
However, when the class of predictive models indexed by $\Theta $ does not
contain the \textit{true} predictive distribution there is no sense in which
this conventional approach remains the `gold standard'. In such cases, the
loss that underpins (\ref{conv_pred}) should be replaced by the \textit{%
	particular} predictive loss that matters for the problem at hand. That is,%
\textbf{\ }our prior beliefs about ${\theta }$ and, hence, about the
elements $P_{\theta }^{(n)}$ in $\mathcal{P}^{(n)}$, need to be updated via
a criterion function defined by a user-specified measure of predictive loss.

\subsection{Bayesian Updating Based on Scoring Rules}

\cite{loaiza2019focused} propose a method for producing Bayesian predictions
using loss functions that specifically capture {the accuracy of} {density
	forecasts}. For $\mathcal{P}^{(n)}$ a convex class of {predictive
	distributions} on $(\Omega ,\mathcal{F})$, density prediction accuracy can
be measured using the positively-oriented\textbf{\ }(i.e. higher is better)
scoring rule $s:\mathcal{P}^{(n)}\times \mathcal{Y}\mapsto \mathbb{R} $,
where the expected scoring rule under the true distribution $P_{0}$ is
defined as 
\begin{equation}
\mathbb{S}(\cdot ,P_{0}):=\int_{y\in \Omega }s(\cdot ,y)\dt P_{0}(y).
\label{exp_score}
\end{equation}%
{Since} $\mathbb{S}(\cdot ,P_{0})$ is unattainable in practice, a sample
estimate based on $y_{1:n}$ is used to define a sample criterion: for a
given $\theta \in \Theta $, define sample average score as 
\begin{equation}
S_{n}(\theta ):=\sum_{t=0}^{n-1}s(P_{\theta }^{(t)},y_{t+1}).
\label{eq:expected_score}
\end{equation}

Adopting the \textit{generalized }updating rule proposed by \cite%
{bissiri2016general} (see also \citealp{giummole2017objective}, %
\citealp{holmes2017assigning}, \citealp{lyddon2019general}, and %
\citealp{Syring2019}), \cite{loaiza2019focused} distinguish elements in $%
\Theta $ using 
\begin{equation}
\pi _{w}(\theta |y_{1:n})=\frac{\exp \left[ wS_{n}\left( \theta \right) %
	\right] \pi (\theta )}{\int_{\Theta }\exp \left[ wS_{n}\left( \theta \right) %
	\right] \pi (\theta )\dt\theta },  \label{post}
\end{equation}%
where the scale factor\textbf{\ }$w$ is obtained in a preliminary step using
measures of predictive accuracy. This posterior explicitly weights elements
of $\Theta $ according to their predictive accuracy in the scoring rule 
$s(\cdot ,\cdot )$. As such, the one-step-ahead 
generalized predictive, 
\begin{equation}
P_{\Pi _{w}}^{(n)}:=\int_{\Theta }P_{\theta }^{(n)}\pi _{w}(\theta |y_{1:n})%
\dt\theta ,  \label{loss_pred}
\end{equation}%
will often outperform, in the chosen rule $s(\cdot ,\cdot )$, the likelihood
(or {log-score})-based predictive $P_{\Pi }^{(n)}$ in cases where the model
is misspecified. Given its explicit dependence {on the Gibbs
	posterior in (\ref{post})}, and as noted earlier, we refer to the predictive
in (\ref{loss_pred}) as the (exact) Gibbs predictive.\footnote{We note that, in general the choice of $w$ can be $n$-dependent. However, we eschew this dependence to maintain notational brevity.}

\section{Gibbs Variational Prediction\label{GVP}}

\subsection{Overview}

While $\Pi _{w}(\cdot |y_{1:n})$ is our \textit{ideal} posterior, it can be
difficult to sample from if the dimension of $\theta $ is {even moderately}
large, which occurs in situations with a large number of predictors, or in
flexible models. Therefore, the exact predictive itself is not readily
available in such cases. However, this does not invalidate the tenant on
which $P_{\Pi _{w}}^{(n)}$ is constructed. Viewed in this way, we see that
the problem of predictive inference via $P_{\Pi _{w}}^{(n)}$ could be solved
if one were able to construct an accurate enough approximation to $\Pi
_{w}(\cdot |y_{1:n})$. Herein, we propose to approximate $P_{\Pi _{w}}^{(n)}$
using variational Bayes (VB).

VB seeks to approximate $\Pi _{w}(\cdot |y_{1:n})$ by finding the closest
member in a class of probability measures, denoted as $\mathcal{Q}$, to $\Pi
_{w}(\cdot |y_{1:n})$ in a chosen divergence measure. The most common choice
of divergence is the Kullback-Leibler (KL) divergence: for probability measures $%
P,Q$, and $P$ absolutely continuous with respect to $Q$, the KL divergence is given by $\text{D}%
(P||Q)=\int \log (\dt P/\dt Q)\dt P$. VB then attempts to produce a
posterior approximation to $\Pi _{w}$ by choosing $Q$ to minimize: 
\begin{equation*}
\text{D}\left( Q||\Pi _{w}\right) =\int \log \left[ {\dt Q}/{\dt\Pi
	_{w}(\cdot |y_{1:n})}\right] \dt Q.
\end{equation*}%

In practice, VB is often conducted, in an equivalent manner, by maximizing the so-called
evidence lower bound (ELBO). In the case where $\Pi_w$ and $Q$ admit densities $\pi_w$ and $q$, respectively, the ELBO is given by: 
\begin{equation}
\text{ELBO}[Q||\Pi _{w}]:=\mathbb{E}_{q}[\log \left\{ \exp \left[
wS_{n}(\theta )\right] \pi (\theta )\right\} ]-\mathbb{E}_{q}[\log
\{q(\theta )\}].  \label{ELBO}
\end{equation}
While other classes of divergences are used in variational inference, such
as the R\'{e}nyi-divergence, the KL divergence is the most commonly
encountered in the literature. Hence, in what follows we focus on this
choice, but note here that other divergences can also be used.

The variational approximation to the Gibbs posterior, $P_{\Pi _{w}}^{(n)}$, can
be defined as follows.

\begin{definition}\label{def:two} For $\mathcal{Q}$ a class of probability distributions, the
	variational posterior $\widehat{Q}$ satisfies 
	\begin{equation*}
	\widehat{Q}:=\operatornamewithlimits{argmin\,}_{Q\in \mathcal{Q}}\text{D}%
	\left[ Q||\Pi _{w}(\cdot |y_{1:n})\right] .
	\end{equation*}%
	and the 
	{Gibbs} variational
	predictive (GVP) is defined as 
	\begin{equation*}
	P_{Q}^{(n)}:=\int_{\Theta }P_{\theta }^{(n)}\dt\widehat{Q}(\theta ).
	\end{equation*}
\end{definition}

\begin{rem}
	In general, we only require that $\widehat{Q}$ is an approximate minimizer,
	i.e., $\text{D}[ \widehat Q||\Pi _{w}(\cdot |y_{1:n})]\ge \sup_{Q\in%
		\mathcal{Q}}\text{D}[ \widehat Q||\Pi _{w}(\cdot |y_{1:n})] +o_p(1)$. The later may
	be useful in cases where $\widehat{Q}$ may not be unique, or in cases where
	it does not exist for a given $n$ but is well-behaved for all $n$ large
	enough.
\end{rem}

The GVP, $P_{Q}^{(n)}$, circumvents the need to construct $%
P_{\Pi _{w}}^{(n)}$ via sampling the Gibbs posterior $\Pi _{w}(\cdot
|y_{1:n}) $. In essence, we replace the sampling problem with an
optimization problem for which reliable methods exist even if $\Theta $ is
high-dimensional, and which in turn yields an approximation to $\Pi
_{w}(\cdot |y_{1:n})$. Consequently, even though, as discussed earlier, it
is not always feasible to {access} $P_{\Pi _{w}}^{(n)}$, via simulation from, 
$\Pi _{w}(\cdot |y_{1:n})$ in situations where $\Theta $ is
high-dimensional, {access to the variational predictive  }$P_{Q}^{(n)}$ 
remains feasible.

This variational approach to prediction is related to the `generalized
variational inference'\ approach of \cite{GVI2019},\textbf{\ }but with two
main differences. Firstly, our approach is focused on \textit{predictive
	accuracy}, not on parameter inference. Our only goal is to produce density
forecasts that are as accurate as possible in the chosen scoring rule $%
s(\cdot ,y)$.\footnote{%
	Critically, since predictive accuracy is our goal, we are not concerned with
	the potential over/under-coverage of credible sets for the model unknowns built from generalized
	posteriors, or VB posteriors; see \cite{miller2021asymptotic}
	for a theoretical discussion of posterior coverage in generalized Bayesian
	inference.} Secondly, our approach follows the idea of \cite%
{bissiri2016general} and \cite{loaiza2019focused} and targets the
predictions built from the Gibbs posterior in \eqref{post}, rather than the
exponentiated form of some general loss as in \cite{GVI2019}. This latter
point is certainly critical if inferential accuracy were still deemed to be
important, since without the tempering constant $w$ that defines the
posterior in \eqref{post}, the exponentiated\ loss function can be very flat
and the posterior $\Pi _{w}(\cdot |y_{1:n})$ uninformative about $\theta $.%
\footnote{%
	See \cite{bissiri2016general}, \cite{giummole2017objective}, \cite%
	{holmes2017assigning}, \cite{lyddon2019general}, \cite{Syring2019} and \cite%
	{pacchiardi2021generalized}{\ for various approaches to the setting of }$w$ {%
		\ in inferential settings.}} It is our experience however (%
\citealp{loaiza2019focused}), that in settings where predictive accuracy is
the only goal, the choice of $w$ actually has little impact on the
generation of accurate predictions via $P_{\Pi _{w}}^{(n)}$, as long as the
 sample size is reasonably large.
Preliminary experimentation in the current setting, in which the variational
predictive $P_{Q}^{(n)}$ is the target, suggests that this finding remains
relevant, and as a consequence we have adopted a default value of $w=1$ in
all numerical work. Further research is needed to deduce the precise impact
of $w$ on $P_{Q}^{(n)}$, in particular for smaller sample sizes, and we
leave this important topic for future work.\footnote{We note here that the work of \cite{wu2021calibrating} attempts to calibrate the coverage of posterior predictives constructed from power posteriors. While the approach of \cite{wu2021calibrating} is not directly applicable in our context, we conjecture that an appropriately modified version of their approach could be applied in our setting to derive posterior predictives with reasonable predictive coverage.}

\subsection{Accuracy of the GVP\label{theory}}

By driving the updating mechanism by the {measure of predictive accuracy}
that matters, it is hoped that {GVP produces} accurate predictions without
requiring correct model specification. The following result demonstrates
that, in large samples, the GVP, $P^{(n)}_Q$, produces predictions that are
indistinguishable from the exact predictive, $P^{(n)}_{\Pi_{w}}$; thus, in
terms of predictive accuracy, the use of a variational approximation to the
Gibbs posterior has little impact on the accuracy of the posterior
predictives. {For probability measures $P,Q$, let $d_{\text{TV} }\{P,Q\}$
	denote the total variation distance. }

\begin{thm}
	\label{thm:merge1} Under Assumptions \ref{ass:est}-\ref{ass:post} in Appendix \ref{app:theory}, $d_{\text{TV} }\{P_{Q}^{(n)},P_{\Pi _{w}}^{(n)}\}\rightarrow 0$ in probability.
\end{thm}

Theorem \ref{thm:merge1} demonstrates that the difference between
distributional predictions made using the GVP and the exact Gibbs predictive agree in the
rule $s(\cdot ,\cdot ) $ in large samples. This type of result is
colloquially known as a `merging' result (\citealp{blackwell1962merging}), and means that if {we take
	the exact Gibbs predictive as our benchmark for (Bayesian) prediction, then
	the GVP is as accurate as this possibly infeasible benchmark (at least in
	large samples).}

\begin{rem}
	As an intermediate step in the proof of Theorem \ref{thm:merge1}, we must
	demonstrate posterior concentration of $\widehat{Q}$, and $\Pi _{w}$, in
	cases where the data is temporally dependent, 
	{possibly} heavy-tailed,
	and where $S_{n}(\theta )$ is an arbitrary scoring rule. For the sake of
	brevity, and to keep the focus on the practical usefulness of this approach,
	we present these details in Appendix \ref{app:theory}. However, we note that our
	focus on the types of data often encountered in forecasting, {for
		example,} financial, economic, or atmospheric processes, does not allow us
	to use existing approaches to guarantee posterior concentration of $\widehat{%
		Q}$ and $\Pi _{w}$. Instead, we must rely on the smoothness of $%
	S_{n}(\theta )$ and the control of certain remainder terms to demonstrate
	this concentration. We refer the interested reader to Appendix \ref{app:theory} for full details and discussion.
\end{rem}

As a further result, we can demonstrate that GVP is as accurate as the infeasible frequentist
`optimal [distributional] predictive' approach suggested in \cite{gneiting2007strictly}.
{Following \cite{gneiting2007strictly}, define the `optimal' (frequentist)
	predictive within the class $\mathcal{P}^{(n)}$, and based on scoring
	rule $s(\cdot ,y)$, as $P_{\star }^{(n)}:=P(\cdot |\mathcal{F}_{n},\theta
	_{\star })$, where } 
\begin{equation}
\theta _{\star }:=\arg \max_{\theta \in \Theta }\mathcal{S}(\theta ),\text{ {%
		with} }\mathcal{S}(\theta )=\operatornamewithlimits{lim\,}_{n\rightarrow
	\infty }\mathbb{E}\left[ S_{n}(\theta )/n\right] .  \label{eq:limit_loss}
\end{equation}%
The following result demonstrates that, in large samples, predictions
produced using {the GVP} are equivalent to those made using the optimal frequentist predictive $%
P_{\star }^{(n)}$.

\begin{thm}
	\label{thm:merge} Under Assumptions \ref{ass:est}-\ref{ass:post} in Appendix \ref{app:theory}, $d_{\text{TV} }\{P_{Q}^{(n)},P_{\star }^{(n)}\}\rightarrow 0$ in probability.
\end{thm}

We remind the reader that, for the sake of brevity, statement of the assumptions, and proofs of all stated results, are given in Appendix \ref{app:theory}.

\section{A Toy Model of Financial Returns\label{sec:toy}}

We now illustrate the behavior of GVP in a simple toy example for a
financial return, $Y_{t}$, in which the exact Gibbs predictive is also
accessible.\footnote{%
	We reiterate here, and without subsequent repetition, that all details of the
	prior specifications and the numerical steps required to produce the
	variational approximations, for this and the following numerical examples,
	are provided in {the supplementary appendices}.} This example serves {two
	purposes. First, it illustrates} {Theorem \ref{thm:merge}}, and, in so doing, highlights the benefits of GVP relative to
prediction based on a misspecified likelihood function. {Secondly}, 
{the }GVP {is shown to }yield almost equal (average)
out-of-sample predictions to those produced by the exact Gibbs posterior
accessed via MCMC;{\ i.e. numerical support for Theorem \ref%
	{thm:merge1} is provided.}

The predictive class, $\mathcal{P}^{(n)}$, is defined by a generalized
autoregressive conditional heteroscedastic GARCH(1,1) model with Gaussian
errors, $Y_{t}=\theta _{1}+\sigma _{t}\varepsilon _{t},$\ $\varepsilon _{t}%
\overset{ i.i.d.}{\sim }N\left( 0,1\right) ,$\ $\sigma _{t}^{2}=\theta
_{2}+\theta _{3}\left( Y_{t-1}-\theta _{1}\right) ^{2}+\theta _{4}\sigma
_{t-1}^{2},$\ with $\theta =\left( \theta _{1},\theta _{2},\theta _{3},{%
	\theta }_{4}\right) ^{\prime }$. We adopt three alternative specifications
for the true data generating process (DGP) : 1) a model that matches the
Gaussian GARCH(1,1) predictive class; 2) a stochastic volatility model with
leverage: 
\begin{align*}
Y_{t}& =\exp \left( \frac{h_{t}}{2}\right) \varepsilon _{t},\quad h_{t}
=-2+0.7\left( h_{t-1}-(-2)\right) +\eta _{t},\quad \left( \varepsilon
_{t},\eta _{t}\right) ^{\prime } \overset{i.i.d.}{\sim }N\left( 0,%
\begin{bmatrix}
1 & -0.35 \\ 
-0.35 & 0.25%
\end{bmatrix}%
\right) ;
\end{align*}%
and 3) a stochastic volatility model with a smooth transition in the
volatility autoregression:%
\begin{align*}
Y_{t}& =\exp \left( \frac{h_{t}}{2}\right) \varepsilon _{t},\quad h_{t}
=0.9g\left( h_{t-1}\right) h_{t-1}+\eta _{t},\quad \eta _{t} \overset{i.i.d.}%
{\sim }N\left( 0,0.25\right) ,
\end{align*}%
where $g(x)=\left( 1+\exp \left( -2x\right) \right) ^{-1}$. DGP 1) defines a%
\textbf{\ }\textit{correct specification}\textbf{\ }setting, whilst DGPs 2)
and 3) characterize different forms of\textbf{\ }\textit{misspecification.}

Denoting the predictive density function associated with the Gaussian GARCH
(1,1) model, evaluated at the observed\textbf{\ }$y_{t+1}$,\textbf{\ }as $%
p_{\theta }(y_{t+1}|\mathcal{F}_{t})$, we implement GVP using four
alternative forms of (postively-oriented) scoring rules:%
\begin{align}
& s^{\text{LS}}\left( P_{\theta }^{(t)},y_{t+1}\right) =\log p_{\theta
}(y_{t+1}|\mathcal{F}_{t}),  \label{ls} \\
& s^{\text{CRPS}}\left( P_{\theta }^{(t)},y_{t+1}\right) =-\int_{-\infty
}^{\infty }\left[ P_{\theta }^{(t)}-I(y\geq y_{t+1})\right] ^{2}dy,
\label{crps} \\
& s^{\text{CLS-A}}\left( P_{\theta }^{(t)},y_{t+1}\right) =\log p_{\theta
}(y_{t+1}|\mathcal{F}_{t})I\left( y_{t+1}\in A\right) +\left[ \ln
\int_{A^{c}}p_{\theta }(y|\mathcal{F}_{t})dy\right] I\left( y_{t+1}\in
A^{c}\right) ,  \label{cls} \\
& s^{\text{IS}}\left( P_{\theta }^{(t)},y_{t+1}\right) =\left(
u_{t+1}-l_{t+1}\right) +\frac{2}{\alpha }\left( l_{t+1}-y_{t+1}\right)
I\left( y_{t+1}<l_{t+1}\right) +\frac{2}{\alpha }\left(
y_{t+1}-u_{t+1}\right) I\left( y_{t+1}>u_{t+1}\right) ,  \label{msis}
\end{align}%
where $l_{t+1}\ $and $u_{t+1}$\ denote the $100(\frac{\alpha }{2})\%$\ and $%
100(1-\frac{\alpha }{2})\%\ $predictive quantiles. The log-score (LS) in (%
\ref{ls}) is a `local' scoring rule, attaining a high value if the observed
value, $y_{t+1}$, is in the high density region of $p_{\theta
}(y_{t+1}|F_{t})$. The continuously ranked probability score (CRPS) in (\ref%
{crps}) (\citealp{gneiting2007strictly}) is, in contrast, sensitive to
distance,\ and rewards the assignment of high predictive mass near to the
realized $y_{t+1}$, rather than just at that value. The score in (\ref{cls})
is the censored likelihood score (CLS)\ of \cite{diks2011likelihood}, which
rewards predictive accuracy over any pre-specified region of interest $A$\ ($%
A^{c}$\ denoting the complement). We use the score for $A$\ defining the
lower and upper tails of the predictive distribution, as determined in turn
by the 10\%, 20\%, 80\% and 90\% percentiles\ of the empirical distribution
of $Y_{t}$, labelling these cases hereafter as CLS$10$, CLS$20,$ CLS$80$ and
CLS$90.$ A high value of this score in any particular instance thus reflects
a predictive distribution that accurately predicts extreme values of the
financial return. The last score considered is
the{\ interval score (IS)} in (\ref{msis}), which is designed to
measure the accuracy of the $100(1-\alpha )\%$\ predictive interval where $%
\alpha =0.05$. This score rewards narrow intervals with accurate coverage.
All components of (\ref{cls}) have closed-form solutions for the
(conditionally) Gaussian predictive model, as does the integral in (\ref%
{crps}) and the bounds in (\ref{msis}).

In total then, seven distinct scoring rules are used to define the sample
criterion function in (\ref{eq:expected_score}). In what follows we
reference these seven criterion functions, and the associated Gibbs
posteriors using the notation\textbf{\ }$S_{n}^{j}\left( \theta \right)
=\sum_{t=0}^{n-1}s^{j}\left( P_{\theta }^{(t)},y_{t+1}\right) $,\textbf{\ }%
for\textbf{\ }$j=\{\text{LS, CRPS, CLS10, CLS20, CLS80, }${CLS90, IS}$%
\}$.

\subsection{Estimation of the Gibbs Predictives}

Given the {low dimension} of the predictive model it is straightforward to
use an MCMC scheme to sample from the exact Gibbs posterior $\pi _{w}^{j}({%
	\theta }|y_{1:n})\propto \exp \{wS_{n}^{j}(\theta )\}\pi \left( \theta
\right) , $ where $\pi _{w}^{j}(\theta |y_{1:n})$ is the exact Gibbs
posterior density in (\ref{post}) computed under scoring rule $j.$ As noted
earlier, in this and all following {numerical} work we set $w=1$. For each
of the $j$ posteriors, we initialize the chains using a burn-in period of $%
20000$ periods, and retain the next $M=20000$\ draws $\theta ^{(m,j)}\sim
\pi _{w}^{j}(\theta |y_{1:n}),$ $m=1,\dots ,M$. The posterior draws are then
used to estimate the exact Gibbs predictive in (\ref{loss_pred}) via $\hat{P}%
_{\Pi _{w}}^{(n,j)}=\frac{1}{M}\sum_{m=1}^{M}P_{\theta ^{(m,j)}}^{(n)}. $

To perform GVP, we first need to produce the variational approximation of $%
\pi _{w}^{j}(\theta |y_{1:n}).$ This is achieved in several steps. First,
the parameters of the GARCH(1,1) model are transformed to the real line.
With some abuse of notation, we re-define here the GARCH(1,1) parameters
introduced in the previous section with the superscript $r$\ to signal
`raw'. The parameter vector $\theta $ is then re-defined as $\theta =\left(
\theta _{1},\theta _{2},\theta _{3},\theta _{4}\right) ^{\prime }=\left(
\theta _{1}^{r},\log (\theta _{2}^{r}),\Phi _{1}^{-1}\left( \theta
_{3}^{r}\right) ,\Phi _{1}^{-1}\left( \theta _{4}^{r}\right) \right)
^{\prime }$, where $\Phi _{1}$\ denotes the {(univariate)} normal cumulative
distribution function (cdf). The next step involves approximating $\pi
_{w}^{j}(\theta |y_{1:n})$ for the re-defined $\theta $. We adopt the
mean-field variational family (see for example \citealp{blei2017variational}%
), with a product-form Gaussian density $q_{\lambda }\left( \theta \right)
=\prod_{i=1}^{4}\phi _{1}\left( {\theta }_{i};\mu _{i},d_{i}^{2}\right) $,
where $\lambda =\left( \mu ^{\prime },d^{\prime }\right) ^{\prime }$ is the
vector of variational parameters, {comprised of }mean and variance vectors $%
\mu =\left( \mu _{1},\mu _{2},\mu _{3},\mu _{4}\right) ^{\prime }$ and $%
d=\left( d_{1},\dots ,d_{4}\right) ^{\prime }$\ respectively, {and }$\phi
_{1}$ {denotes the {(univariate) }normal probability density function (pdf). 
}Denote as $Q_{\lambda }$ and $\Pi _{w}^{j}$ the {distribution functions
	associated with} $q_{\lambda }\left( \theta \right) $ and $\pi
_{w}^{j}(\theta |y_{1:n})$, respectively. The approximation is then
calibrated by solving the maximization problem 
\begin{equation}
\tilde{{\lambda }}=\text{ }\underset{\lambda \in \Lambda }{\arg \max }\ 
\mathcal{L}(\lambda ),  \label{Eq:optimization}
\end{equation}%
where, $\mathcal{L}(\lambda ):=\text{ELBO}\left[ Q_{\lambda }||\Pi _{w}^{j}%
\right] $. Remembering the use of the notation $S_{n}^{j}\left( \theta
\right) $ to denote (\ref{eq:expected_score}) for scoring rule $j$, (\ref%
{ELBO}) (for case $j$) becomes $\mathcal{L}(\lambda )=\mathbb{E}_{q_{\lambda
}}\left[ wS_{n}^{j}\left( {\theta }\right) +\log \pi \left( {\theta }\right)
-\log q_{\lambda }\left( {\theta }\right) \right] .$ Optimization is
performed via SGA, as described in
Section \ref{sga} of the supplementary appendix. Once calibration of $\tilde{\lambda}$ is completed, the
GVP is estimated as $\hat{P}_{Q}^{(n,j)}=\frac{1}{M}\sum_{m=1}^{M}P_{\theta
	^{(m,j)}}^{(n)}$ with $\theta ^{(m,j)}\overset{i.i.d.}{\sim }q_{\tilde{%
		\lambda}}\left( {\theta }\right) $. To calibrate $\tilde{\lambda}$ $10000$
VB iterations are used, and to estimate the variational predictive we set $%
M=1000$.

To produce the numerical results we generate a times series of length $T=6000
$ from the true DGP. Then, we {perform} an expanding window exercise
from $n=1000$ to $n=5999$. For $j\in \{\text{LS, CRPS, CLS10, CLS20, CLS80, CLS90, IS\}}$ we construct the predictive densities $\hat{P}_{\Pi _{w}}^{(n,j)}
$ and $\hat{P}_{Q}^{(n,j)}$ as outlined\ above. Then, for $i\in \{\text{LS,
	CRPS, CLS10, CLS20, CLS80, CLS90, IS\}}$ we compute the measures of
out-of-sample predictive accuracy $S_{\Pi _{w}}^{i,j,n}=s^{i}\left( \hat{P}%
_{\Pi _{w}}^{(n,j)},y_{n+1}\right) $ and $S_{Q}^{i,j,n}=s^{i}\left( \hat{P}%
_{Q}^{(n,j)},y_{n+1}\right) $. Finally, we compute the average out-of-sample
scores $S_{\Pi _{w}}^{i,j}=\frac{1}{5000}\sum_{n=1000}^{5999}S_{\Pi
	_{w}}^{i,j,n}$ and $S_{Q}^{i,j}=\frac{1}{5000}%
\sum_{n=1000}^{5999}S_{Q}^{i,j,n}$. The results are tabulated and discussed
in the following section.

\subsection{Results}

The results of the simulation exercise are recorded in Table~\ref%
{tab:toyExampleGARCH(1,1)}. Panels A to C record the results for Scenarios
1) to 3), with the average out-of-sample scores associated with the exact
Gibbs predictive (estimated via MCMC) appearing in the left-hand-side panel
and the average scores for GVP appearing in the right-hand-side panel. All
values on the diagonal of each sub-panel correspond to the case where $i=j$.
In the misspecified case, numerical validation of the asymptotic result that
the GVP concentrates onto the optimal predictive ({Theorem \ref%
	{thm:merge}}) occurs if the largest values (bolded) in a column appear on
the diagonal (`strict coherence' in the language of %
\citealp{martin2020optimal}). In the correctly specified case, in which all
proper scoring rules will, for a large enough sample, pick up {\ the one}
true model (\citealp{gneiting2007strictly}), we would expect all values in
given column to be very similar to one another, differences reflecting
sampling variation only. Finally, validation of the theoretical property of
merging between the exact and variational Gibbs predictive ({Theorem %
	\ref{thm:merge1}}) occurs if the corresponding results in all left- and
right-hand-side panels are equivalent.

As is clear, there is almost uniform numerical validation of {both
	theoretical results}. Under misspecification Scenario 2), (Panel B) the GVP
results are strictly coherent (i.e. all bold values lie on the diagonal),
with the exact Gibbs predictive results equivalent to the corresponding GVP
results to three or four decimal places. The same broad findings obtain
under misspecification Scenario 3), apart from the fact that the {IS
	updates} are second best (to the log-score; and then only just) in terms of
the out-of-sample {IS measure}. In Panel A on the other hand, we see
the expected (virtual) equivalence of all results in a given column,
reflecting the fact that all Gibbs predictives (however estimated) are
concentrating onto the true predictive model and, hence, have identical
out-of-sample performance. Of course, for a large but still finite number of
observations, we would expect the log-score to perform best, due to the
efficiency of the implicit maximum likelihood estimator underpinning the
results and, to all intents and purposes this is exactly what we observe in
Panels A.1 and A.2.

In summary, GVP performs as anticipated, and reaps distinct benefits in
terms of predictive accuracy. Any inaccuracy in the measurement of parameter
uncertainty also has negligible impact on the finite sample performance of
GVP relative to an exact comparator. {In Section \ref{complex}}, we
extend the investigation into design settings that mimic the
high-dimensional problems to which we would apply the variational approach
in practice, followed by an empirical application - again using a
high-dimensional predictive model - in Section \ref{emp}.

\clearpage

\begin{sidewaystable}[ph!]\centering
	\scalebox{0.85}{
		\begin{tabular}{lcccccccrccccccc}
			\toprule\toprule
			&                                                                                                                                             \multicolumn{15}{c}{\textbf{Panel A. True DGP:  GARCH(1,1)}}                                                                                                                                             \\
			&                                               \multicolumn{7}{c}{\textbf{A.1: Exact Gibbs predictive}}                                                &  &                                                                             \multicolumn{7}{c}{\textbf{A.2: GVP}}                                                                              \\
			&                                          \multicolumn{7}{c}{\textbf{Average out-of-sample score}}                                          &  &                                                                       \multicolumn{7}{c}{\textbf{Average out-of-sample score}}                                                                       \\
			\cmidrule{2-8}\cmidrule{10-16} &                  &                  &                  &                  &                  &                  &                  &  &                        &                           &                           &                           &                           &                          &                          \\
			&        LS        &      CLS10       &      CLS20       &      CLS80       &      CLS90       &       CRPS       &       IS       &  & \multicolumn{1}{l}{LS} & \multicolumn{1}{l}{CLS10} & \multicolumn{1}{l}{CLS20} & \multicolumn{1}{l}{CLS80} & \multicolumn{1}{l}{CLS90} & \multicolumn{1}{l}{CRPS} & \multicolumn{1}{l}{IS} \\
			U.method                       &                  &                  &                  &                  &                  &                  &                  &  &                        &                           &                           &                           &                           &                          &                          \\
			\cmidrule{1-1}    LS           & \textbf{-0.0433} & \textbf{-0.2214} & \textbf{-0.3087} & \textbf{-0.3029} &     -0.2150      & \textbf{-0.1438} & \textbf{-1.1853} &  &    \textbf{-0.0434}    &     \textbf{-0.2216}      &     \textbf{-0.3088}      &     \textbf{-0.3028}      &     \textbf{-0.2150}      &     \textbf{-0.1438}     &     \textbf{-1.1857}     \\
			CLS10                          &     -0.0485      &     -0.2222      &     -0.3094      &     -0.3069      &     -0.2190      &     -0.1441      &     -1.2009      &  &        -0.0496         &          -0.2227          &          -0.3100          &          -0.3075          &          -0.2199          &         -0.1441          &         -1.2053          \\
			CLS20                          &     -0.0482      &     -0.2221      &     -0.3094      &     -0.3066      &     -0.2185      &     -0.1440      &     -1.2014      &  &        -0.0493         &          -0.2225          &          -0.3097          &          -0.3072          &          -0.2193          &         -0.1440          &         -1.2052          \\
			CLS80                          &     -0.0558      &     -0.2305      &     -0.3198      &     -0.3032      &     -0.2150      &     -0.1446      &     -1.2087      &  &        -0.0567         &          -0.2311          &          -0.3204          &          -0.3033          &          -0.2152          &         -0.1446          &         -1.2107          \\
			CLS90                          &     -0.0495      &     -0.2258      &     -0.3143      & \textbf{-0.3029} &     -0.2150      &     -0.1441      &     -1.1973      &  &        -0.0502         &          -0.2265          &          -0.3149          &          -0.3031          &          -0.2152          &         -0.1441          &         -1.1993          \\
			CRPS                           &     -0.0462      &     -0.2239      &     -0.3116      &     -0.3027      & \textbf{-0.2147} & \textbf{-0.1438} &     -1.1874      &  &        -0.0474         &          -0.2239          &          -0.3112          &          -0.3041          &          -0.2162          &         -0.1439          &         -1.1918          \\
			IS                           &     -0.0438      &     -0.2216      &     -0.3089      &     -0.3031      &     -0.2153      & \textbf{-0.1438} &     -1.1877      &  &        -0.0440         &          -0.2218          &          -0.3091          &          -0.3031          &          -0.2153          &     \textbf{-0.1438}     &         -1.1882          \\ \toprule\toprule
			&                                                                                                                                       \multicolumn{15}{c}{\textbf{Panel B. True DGP: Stochastic volatility with leverage}}                                                                                                                                        \\
			&                                               \multicolumn{7}{c}{\textbf{B.1: Exact Gibbs predictive}}                                                &  &                                                                             \multicolumn{7}{c}{\textbf{B.2: GVP}}                                                                              \\
			&                                          \multicolumn{7}{c}{\textbf{Average out-of-sample score}}                                          &  &                                                                       \multicolumn{7}{c}{\textbf{Average out-of-sample score}}                                                                       \\
			\cmidrule{2-8}\cmidrule{10-16} &                  &                  &                  &                  &                  &                  &                  &  &                        &                           &                           &                           &                           &                          &                          \\
			&        LS        &      CLS10       &      CLS20       &      CLS80       &      CLS90       &       CRPS       &       IS       &  & \multicolumn{1}{l}{LS} & \multicolumn{1}{l}{CLS10} & \multicolumn{1}{l}{CLS20} & \multicolumn{1}{l}{CLS80} & \multicolumn{1}{l}{CLS90} & \multicolumn{1}{l}{CRPS} & \multicolumn{1}{l}{IS} \\
			U.method                       &                  &                  &                  &                  &                  &                  &                  &  &                        &                           &                           &                           &                           &                          &                          \\
			\cmidrule{1-1}    LS           & \textbf{-0.5636} &     -0.3753      &     -0.5452      &     -0.3536      &     -0.2512      &     -0.2313      &     -2.3468      &  &    \textbf{-0.5633}    &          -0.3752          &          -0.545           &          -0.3535          &          -0.2511          &         -0.2313          &         -2.3467          \\
			CLS10                          &     -1.0193      & \textbf{-0.3336} &     -0.5021      &     -0.8379      &     -0.7339      &     -0.3679      &      -3.447      &  &        -1.0156         &     \textbf{-0.3336}      &          -0.502           &          -0.834           &          -0.7302          &         -0.3659          &         -3.4207          \\
			CLS20                          &      -0.806      &     -0.3354      & \textbf{-0.4968} &     -0.6291      &     -0.5267      &     -0.2863      &     -2.9923      &  &        -0.8055         &          -0.3355          &     \textbf{-0.4969}      &          -0.6282          &          -0.5259          &         -0.2861          &         -2.9853          \\
			CLS80                          &     -0.9203      &     -0.7372      &     -0.9311      & \textbf{-0.329}  & \textbf{-0.2292} &     -0.2402      &     -3.3135      &  &        -0.9357         &          -0.7514          &          -0.9463          &      \textbf{-0.329}      &          -0.2292          &         -0.2402          &         -3.3248          \\
			CLS90                          &     -0.9575      &     -0.7615      &     -0.9649      &     -0.3294      & \textbf{-0.2292} &     -0.2425      &     -3.4213      &  &        -0.9959         &          -0.7969          &          -1.0033          &          -0.3293          &     \textbf{-0.2291}      &         -0.2426          &         -3.4476          \\
			CRPS                           &     -0.5692      &     -0.4029      &     -0.5671      &     -0.3431      &     -0.2419      &  \textbf{-0.23}  &     -2.4312      &  &        -0.5649         &          -0.3985          &          -0.5626          &          -0.3432          &          -0.2419          &     \textbf{-0.2301}     &         -2.4338          \\
			IS                           &     -0.6552      &     -0.3986      &     -0.6111      &     -0.3713      &      -0.248      &     -0.2604      & \textbf{-2.203}  &  &         -0.655         &          -0.3985          &          -0.6109          &          -0.3712          &          -0.2479          &         -0.2603          &     \textbf{-2.2033}     \\ \toprule\toprule
			&                                                                                                                                   \multicolumn{15}{c}{\textbf{Panel C. True DGP: Stochastic volatility with smooth transition}}                                                                                                                                   \\
			&                                               \multicolumn{7}{c}{\textbf{C.1: Exact Gibbs predictive}}                                                &  &                                                                             \multicolumn{7}{c}{\textbf{C.2: GVP}}                                                                              \\
			&                                          \multicolumn{7}{c}{\textbf{Average out-of-sample score}}                                          &  &                                                                       \multicolumn{7}{c}{\textbf{Average out-of-sample score}}                                                                       \\
			\cmidrule{2-8}\cmidrule{10-16} &                  &                  &                  &                  &                  &                  &                  &  &                        &                           &                           &                           &                           &                          &                          \\
			&        LS        &      CLS10       &      CLS20       &      CLS80       &      CLS90       &       CRPS       &       IS       &  &           LS           &           CLS10           &           CLS20           &           CLS80           &           CLS90           &           CRPS           &           IS           \\
			U.method                       &                  &                  &                  &                  &                  &                  &                  &  &                        &                           &                           &                           &                           &                          &                          \\
			\cmidrule{1-1}    LS           & \textbf{-1.6858} &     -0.4239      &      -0.672      &     -0.6751      &     -0.4369      &     -0.7196      & \textbf{-7.0726} &  &    \textbf{-1.686}     &          -0.424           &          -0.6721          &          -0.6752          &          -0.4371          &         -0.7196          &     \textbf{-7.0742}     \\
			CLS10                          &     -1.8173      & \textbf{-0.4172} &     -0.6707      &     -0.8016      &     -0.5468      &      -0.821      &     -7.8217      &  &        -1.8145         &     \textbf{-0.4172}      &          -0.6706          &          -0.7987          &          -0.5437          &         -0.8183          &         -7.7717          \\
			CLS20                          &     -1.7173      &      -0.419      & \textbf{-0.6674} &     -0.7067      &     -0.4616      &     -0.7425      &     -7.2516      &  &        -1.7171         &          -0.4192          &     \textbf{-0.6676}      &          -0.7063          &          -0.4611          &          -0.742          &         -7.2481          \\
			CLS80                          &      -1.733      &      -0.465      &     -0.7199      & \textbf{-0.6681} &     -0.4304      &     -0.7511      &     -7.3087      &  &        -1.7339         &          -0.4654          &          -0.7204          &     \textbf{-0.6684}      &          -0.4307          &         -0.7513          &         -7.3159          \\
			CLS90                          &     -1.8777      &     -0.5927      &     -0.8579      &     -0.6731      & \textbf{-0.4288} &     -0.8707      &     -8.1299      &  &        -1.8819         &          -0.5952          &          -0.8608          &          -0.674           &     \textbf{-0.4295}      &         -0.8734          &         -8.1211          \\
			CRPS                           &     -1.6938      &     -0.4283      &     -0.6762      &     -0.6812      &     -0.4435      & \textbf{-0.7184} &     -7.2059      &  &        -1.6938         &          -0.4286          &          -0.6765          &          -0.6809          &          -0.4432          &     \textbf{-0.7185}     &         -7.2121          \\
			IS                           &     -1.6879      &     -0.4244      &     -0.6726      &     -0.6754      &     -0.4374      &     -0.7208      &     -7.0915      &  &        -1.6881         &          -0.4245          &          -0.6728          &          -0.6754          &          -0.4374          &         -0.7208          &          -7.093          \\ \bottomrule\bottomrule
	\end{tabular}	}
	\caption{{Predictive accuracy of GVP using a Gaussian
		GARCH(1,1) predictive model for a financial return. Panel A corresponds to
		the correctly specified case, and Panels B and C to the two different
		misspecified settings as described in the text.}{\protect\small \ } 
	{{The rows in each panel refer to the update method
			(U.method) used. The columns refer to the out-of-sample measure used to
			compute the average scores. The figures in bold are the largest average
			scores according to a given out-of-sample measure.}}}\label{tab:toyExampleGARCH(1,1)}
\end{sidewaystable}

\clearpage

\section{Complex Time Series Examples\label{complex}}

In this section we demonstrate the application of GVP in two realistic
simulated examples. In both cases the assumed predictive model is
high-dimensional and the exact Gibbs posterior, even if accessible in
principle via MCMC, is challenging from a computational point of view. The
mean-field class is adopted in both cases to produce variational
approximations to the Gibbs posterior. The simulation design for each
example (including the choice of $w=1$) mimics that for the toy example,
apart from the obvious changes {made to} the true DGP and the assumed
predictive model, some changes {in} the size of the estimation and
evaluation periods, plus the absence of comparative exact results. For
reasons of computational burden we remove the CRPS update from consideration
in the first example.

\subsection{Example 1: Autoregressive Mixture Predictive Model\label{ar}}

In this example we adopt a true DGP in which $Y_{t}$ evolves according to
the logistic smooth transition autoregressive (LSTAR) process proposed in 
\cite{terasvirta1994specification}: 
\begin{equation}
Y_{t}=\rho _{1}Y_{t-1}+\rho _{2}\left\{ \frac{1}{1+\exp \left[ -\gamma
	\left( Y_{t-1}-c\right) \right] }\right\} y_{t-1}+\sigma _{\varepsilon
}\varepsilon _{t},  \label{ar_mixture}
\end{equation}%
where $\varepsilon _{t}\overset{i.i.d.}{\sim}t_{\nu }$, and $t_{\nu }$
denotes the standardised Student-t distribution with $\nu $ degrees of
freedom. This model has the ability to produce data that exhibits a range of
complex features. For example, it not only allows for skewness in the
marginal density of $Y_{t}$, but can also produce temporal dependence
structures that are asymmetric. We thus use this model as an illustration of
a complex DGP whose characteristics are hard to replicate with simple
parsimoneous models. Hence the need to adopt a highly parameterized
predictive model; plus the need to acknowledge that even that representation
will be misspecified.

The assumed predictive model is based on the flexible Bayesian
non-parametric structure proposed in \cite{antoniano2016nonparametric}. The
predictive distribution for $Y_{t}$, conditional on the observed $y_{t-1}$,
is constructed from a mixture of $K=20$\ Gaussian autoregressive (AR) models
of order one as follows:%
\begin{equation}
P_{\theta }^{(t-1)}=\sum_{k=1}^{K}\tau_{k,t}\Phi _{1}\left[ Y_{t}-\mu ;\beta
_{k,0}+\beta _{k,1}(y_{t-1}-\mu ),\sigma _{k}^{2}\right] ,
\label{Eq:misturepredclass}
\end{equation}%
with time-varying mixture\textbf{\ }weights 
\begin{equation*}
\tau_{k,t}=\frac{\tau_{k}\phi _{1}\left( y_{t-1}-\mu ;\mu
	_{k},s_{k}^{2}\right) }{\sum_{j=1}^{K}\tau_{j}\phi _{1}\left( y_{t-1}-\mu
	;\mu _{j},s_{j}^{2}\right) },
\end{equation*}%
where $\mu _{k}=\frac{\beta _{k,0}}{1-\beta _{k,1}}$ and $s_{k}^{2}=\frac{%
	\sigma _{k}^{2}}{1-\beta _{k,1}^{2}}$. Denoting $\tau=\left( \tau_{1},\dots
,\tau_{K}\right) ^{\prime }$, $\beta _{0}=\left( \beta _{1,0},\dots ,\beta
_{K,0}\right) ^{\prime }$, $\beta _{1}=\left( \beta _{1,1},\dots ,\beta
_{K,1}\right) ^{\prime }$ and $\sigma =\left( \sigma _{1},\dots ,\sigma
_{K}\right) ^{\prime }$, then the full vector of unknown\textbf{\ }%
parameters is $\theta =\left( \mu ,\tau^{\prime },\beta
_{0}^{^{\prime}},\beta _{1}^{\prime },\right.$ $\left.\sigma ^{\prime
}\right) ^{\prime }$, which comprises $1+\left( 4\times 20\right) =81$
elements. GVP is a natural and convenient alternative to exact Gibbs
prediction in this case.

\subsection{Example 2: Bayesian Neural Network Predictive Model\label{nn}}

\label{sect:BNN}

In the second example we consider a true DGP in which the dependent variable 
$Y_{t}$ has a complex non-linear relationship with a set of covariates.
Specifically, the time series process $\{Y_{t}\}_{t=1}^{T}$, is determined
by a three-dimensional stochastic process $\{X_{t}\}_{t=1}^{T}$, with $%
X_{t}=\left( X_{1,t},X_{2,t},X_{3,t}\right) ^{\prime }$. The first two
covariates are jointly distributed as $(X_{1,t},X_{2,t})^{\prime }\overset{%
	i.i.d.}{\sim }N\left( 0,\Sigma \right) $. The third covariate $X_{3,t}$,
independent of the former two, is distributed according to an AR(4) process
so that $X_{3,t}=\sum_{i=1}^{4}\alpha _{i}X_{3,t-i}+\sigma \varepsilon _{t}$%
, with $\varepsilon _{t}\overset{i.i.d.}{\sim }N\left( 0,1\right) $. The
variable $Y_{t}$ is then given by $Y_{t}=X_{t}^{\prime }\beta _{t}$, where $%
\beta _{t}=\left( \beta _{1,t},\beta _{2,t},\beta _{3,t}\right) ^{\prime }$
is a three dimensional vector of time-varying coefficients, with $\beta
_{i,t}=b_{i}+a_{i}F\left( X_{3,t}\right) $, and $F$ denotes the marginal
distribution function of $X_{3,t}$ induced by the AR(4) model.

This particular choice of DGP has two advantages. First, given the complex
dependence structure of the DGP (i.e. non-linear cross-sectional dependence
as well as temporal dependence), it would be difficult, once again, to find
a simple parsimoneous predictive model that would adequately capture this
structure; hence motivating the need to adopt a high-dimensional model and
to resort to GVP. Second, because it includes several covariates, we can
assess how GVP performs for varying informations sets. For example, we can
establish if the overall performance of GVP improves with an expansion of
the conditioning information set, as we would anticipate; and if the
inclusion of a more complete conditioning set affects the occurrence of
strict coherence, or not.

A flexible model that has the potential to at least partially capture
several features of the true DGP\textbf{\ }is a Bayesian feed-forward neural
network that takes $Y_{t}$ as the dependent variable and some of its lags,
along with observed values of\textbf{\ }$X_{t}$, $x_{t}=\left(
x_{1,t},x_{2,t},x_{3,t}\right) ^{\prime },$ as the vector of\textbf{\ }%
independent inputs, $z_{t}$ (see for instance %
\citealp{hernandez2015probabilistic}). The predictive distribution is
defined as $P_{\theta }^{(t-1)}=\Phi _{1}\left( Y_{t};g\left( z_{t};\omega
\right) ,\sigma _{y}^{2}\right) .$ The mean function $g\left( z_{t};\omega
\right) $ denotes a feed-forward neural network with $q=2$ layers, $r=3$
nodes in each layer, a $p$-dimensional input vector $z_{t}$, and a $d$%
-dimensional parameter vector $\omega $ with $d=r^{2}(q-1)+r(p+q+1)+1$. It
allows for a flexible non-linear relationship between $Y_{t}$ and the vector
of observed covariates $z_{t}$. Defining $c=\log (\sigma _{y})$, the full
parameter vector of this predictive class, $\theta =\left( \omega ^{\prime
},c\right) ^{\prime }$, is of dimension $d+1$.

\subsection{Simulation Results}

\subsubsection{Example 1}

A time series of length $T=2500$ for $Y_{t}$ is generated from the LSTAR
model in \eqref{ar_mixture}, with the true parameters set as: $\rho _{1}=0$, 
$\rho _{2}=0.9$, $\gamma =5$, $c=0$, $\sigma _{\varepsilon }=1$ and $\nu =3$%
. With exception of the CRPS, which cannot be evaluated in closed-form for
the mixture predictive class and is not included in the exercise as a
consequence, the same scores in the toy example are considered. With
reference to the simulation steps given earlier, the initial estimation
window is $500$ observations; hence the predictive accuracy results are
based on an evaluation sample of size $2000.$

\clearpage

\begin{table}[h!]
\centering
	\begin{tabular}{lrrrrrr}
		\hline\hline
		&  &  &  &  &  &  \\ 
		& \multicolumn{6}{c}{\textbf{Average out-of-sample score}} \\ \cline{2-7}
		&  &  &  &  &  &  \\ 
		& LS & CLS10 & CLS20 & CLS80 & CLS90 & IS \\ 
		\textbf{U.method} &  &  &  &  &  &  \\ \cline{1-1}
		&  &  &  &  &  &  \\ 
		LS & \textbf{-1.253} & -0.345 & -0.497 & -0.452 & -0.263 & -5.589 \\ 
		CLS10 & -1.445 & -0.346 & -0.512 & -0.555 & -0.321 & -7.279 \\ 
		CLS20 & -1.445 & \textbf{-0.344} & \textbf{-0.496} & -0.589 & -0.349 & -6.674
		\\ 
		CLS80 & -1.333 & -0.414 & -0.571 & \textbf{-0.450} & -0.260 & -5.831 \\ 
		CLS90 & -1.330 & -0.407 & -0.564 & -0.451 & \textbf{-0.259} & -5.730 \\ 
		IS & -1.410 & -0.401 & -0.558 & -0.474 & -0.282 & \textbf{-5.550} \\ 
		\hline\hline
	\end{tabular}%
	\caption{{Predictive accuracy of GVP using a
		autoregressive mixture model for data generated from an LSTAR model. {The
			rows in each panel refer to the update method (U.method) used. The columns
			refer to the out-of-sample measure used to compute the average scores. The
			figures in bold are the largest average scores according to a given
			out-of-sample measure.}}}
			\label{tab:LSTARresults}
\end{table}

The results in Table \ref{tab:LSTARresults} are clear-cut. With one
exception (that of CLS10), the best out-of-sample performance, according to
a given measure, is produced by the version of GVP based on that same
scoring rule. That is, the GVP results are almost uniformly strictly
coherent: a matching of the update rule with the out-of-sample measure
produces the best predictive accuracy in that measure, almost always.

\subsubsection{Example 2}

In this case, we generate a time series of length $T=4000$ from the model
discussed in Section \ref{sect:BNN}, with specifications: $\Sigma _{11}=1$, $%
\Sigma _{22}=1.25$, $\Sigma _{12}=\Sigma _{21}=0.5$, $\sigma ^{2}=0.2$, $%
\alpha _{1}=0.5$, $\alpha _{2}=0.2$, $\alpha _{3}=0.15$, $\alpha _{4}=0.1$, $%
a_{1}=1.3$, $b_{1}=0$, $a_{2}=-2.6$, $b_{2}=1.3$, $a_{3}=-1.5$ and $%
b_{3}=1.5 $. These settings generate data with a negatively skewed empirical
distribution, a non-linear relationship between the observations on $Y_{t}$
and $X_{3,t}$, and autoregressive behavior in $Y_{t}$.

To assess if the performance of GVP is affected by varying information sets,
we consider four alternative specifications for the input vector $z_{t}$ in
the assumed predictive model. These four specifications (labelled as Model
1, Model 2, Model 3 and Model 4, respectively) are: $z_{t}=y_{t-1}$, $%
z_{t}=\left( y_{t-1},x_{1,t}\right) ^{\prime }$, $z_{t}=\left(
y_{t-1},x_{2,t}\right) ^{\prime }$ and $z_{t}=\left(
y_{t-1},x_{1,t},x_{2,t}\right) ^{\prime }$. The dimension of the parameter
vector for each of these model specifications is $d+1=23$, $d+1=26$, $d+1=26$
and $d+1=29$, respectively. Given that the assumed predictive class is
Gaussian, all scoring rules used in the seven updates, and for all four
model specifications, can be evaluated in closed form. Referencing the
simulation steps given earlier, the initial estimation window is $2000$
observations; hence the predictive accuracy results are based on an
evaluation sample of size $2000$ as in the previous example. 

\clearpage

\begin{sidewaystable}[tbph]
\centering
	\resizebox{22.3cm}{!}
	{
		\begin{tabular}{lrrrrrrrrlrrrrrrr}
			\hline\hline
			&                 &                 &                 &                 &                 &                 &                 &  &                   &                 &                 &                 &                 &                         &                 &                 \\
			&            \multicolumn{5}{c}{\textbf{Panel A: Model 1 (${z}_t = y_{t-1}$)}}            &                 &                 &  &                   &     \multicolumn{5}{c}{\textbf{Panel B: Model 2 (${z}_t = \left(y_{t-1},x_{1,t}\right)'$)}}     &                 &                 \\
			&                 &                 &                 &                 &                 &                 &                 &  &                   &                 &                 &                 &                 &                         &                 &                 \\
			&                                  \multicolumn{7}{c}{\textbf{Average out-of-sample score}}                                   &  &                   &                                      \multicolumn{7}{c}{\textbf{Average out-of-sample score}}                                       \\ \cline{2-8}\cline{11-17}
			&                 &                 &                 &                 &                 &                 &                 &  &                   &                 &                 &                 &                 &                         &                 &                 \\
			&              LS &           CLS10 &           CLS20 &           CLS80 &           CLS90 &            CRPS &            IS &  &                   &              LS &           CLS10 &           CLS20 &           CLS80 &                   CLS90 &            CRPS &            IS \\
			\textbf{U.method} &                 &                 &                 &                 &                 &                 &                 &  & \textbf{U.method} &                 &                 &                 &                 &                         &                 &                 \\ \cline{1-1}\cline{10-10}
			&                 &                 &                 &                 &                 &                 &                 &  &                   &                 &                 &                 &                 &                         &                 &                 \\
			LS                &          -1.607 &          -0.489 &          -0.761 &          -0.520 &          -0.310 & \textbf{-0.659} &          -6.579 &  & LS                & \textbf{-1.451} &          -0.494 &          -0.733 &          -0.424 &                  -0.254 & \textbf{-0.550} &          -6.059 \\
			CLS10             &          -1.914 & \textbf{-0.460} &          -0.738 &          -0.852 &          -0.617 &          -0.914 &          -8.174 &  & CLS10             &          -1.858 & \textbf{-0.435} &          -0.686 &          -0.849 &                  -0.626 &          -0.866 &          -7.707 \\
			CLS20             &          -1.766 & \textbf{-0.460} & \textbf{-0.732} &          -0.704 &          -0.478 &          -0.772 &          -7.448 &  & CLS20             &          -1.683 &          -0.438 & \textbf{-0.679} &          -0.686 &                  -0.476 &          -0.708 &          -6.963 \\
			CLS80             &          -1.637 &          -0.514 &          -0.795 & \textbf{-0.514} & \textbf{-0.303} &          -0.678 &          -6.822 &  & CLS80             &          -1.551 &          -0.595 &          -0.850 & \textbf{-0.403} &         \textbf{-0.241} &          -0.587 &          -6.878 \\
			CLS90             &          -1.686 &          -0.534 &          -0.831 &          -0.521 &          -0.307 &          -0.718 &          -6.912 &  & CLS90             &          -1.539 &          -0.540 &          -0.811 &          -0.406 &         \textbf{-0.241} &          -0.605 &          -6.369 \\
			CRPS              &          -1.620 &          -0.513 &          -0.783 & \textbf{-0.514} &          -0.304 &          -0.660 &          -6.752 &  & CRPS              &          -1.510 &          -0.585 &          -0.822 &          -0.411 &                  -0.244 &          -0.553 &          -6.649 \\
			IS              & \textbf{-1.601} &          -0.476 &          -0.753 &          -0.516 &          -0.307 &          -0.660 & \textbf{-6.338} &  & IS              &          -1.452 &          -0.473 &          -0.728 &          -0.414 &                  -0.248 &           0.562 & \textbf{-5.671} \\ \hline\hline
			&                 &                 &                 &                 &                 &                 &                 &  &                   &                 &                 &                 &                 &                         &                 &                 \\
			& \multicolumn{5}{c}{\textbf{Panel C: Model 3 (${z}_t = \left(y_{t-1},x_{2,t}\right)'$)}} &                 &                 &  &                   & \multicolumn{5}{c}{\textbf{Panel D: Model 4 (${z}_t = \left(y_{t-1},x_{1,t},x_{2,t}\right)'$)}} &                 &                 \\
			&                 &                 &                 &                 &                 &                 &                 &  &                   &                 &                 &                 &                 &                         &                 &                 \\
			&                                  \multicolumn{7}{c}{\textbf{Average out-of-sample score}}                                   &  &                   &                                      \multicolumn{7}{c}{\textbf{Average out-of-sample score}}                                       \\ \cline{2-8}\cline{11-17}
			&                 &                 &                 &                 &                 &                 &                 &  &                   &                 &                 &                 &                 &                         &                 &                 \\
			&              LS &           CLS10 &           CLS20 &           CLS80 &           CLS90 &            CRPS &            IS &  &                   &              LS &           CLS10 &           CLS20 &           CLS80 &                   CLS90 &            CRPS &            IS \\
			\textbf{U.method} &                 &                 &                 &                 &                 &                 &                 &  & \textbf{U.method} &                 &                 &                 &                 &                         &                 &                 \\ \cline{1-1}\cline{10-10}
			&                 &                 &                 &                 &                 &                 &                 &  &                   &                 &                 &                 &                 &                         &                 &                 \\
			LS                &          -1.536 &          -0.428 &          -0.688 &          -0.521 &          -0.314 &          -0.617 & \textbf{-5.975} &  & LS                & \textbf{-1.390} &          -0.483 &          -0.716 &          -0.389 &                  -0.229 & \textbf{-0.511} &          -5.890 \\
			CLS10             &          -1.775 & \textbf{-0.396} &          -0.658 &          -0.793 &          -0.555 &          -0.805 &          -6.704 &  & CLS10             &          -1.758 & \textbf{-0.394} &          -0.650 &          -0.784 &                  -0.548 &          -0.790 &          -6.659 \\
			CLS20             &          -1.646 &          -0.397 & \textbf{-0.654} &          -0.664 &          -0.439 &          -0.694 &          -6.277 &  & CLS20             &          -1.656 & \textbf{-0.394} & \textbf{-0.621} &          -0.732 &                  -0.515 &          -0.712 &          -6.236 \\
			CLS80             &          -1.698 &          -0.566 &          -0.858 &          -0.520 & \textbf{-0.301} &          -0.714 &          -7.282 &  & CLS80             &          -2.127 &          -1.049 &          -1.438 & \textbf{-0.356} &                  -0.196 &          -0.736 &          -9.798 \\
			CLS90             &          -1.744 &          -0.581 &          -0.889 &          -0.530 &          -0.304 &          -0.758 &          -7.361 &  & CLS90             &          -2.157 &          -0.989 &          -1.401 &          -0.379 &         \textbf{-0.192} &          -0.782 &          -9.545 \\
			CRPS              & \textbf{-1.535} &          -0.438 &          -0.695 & \textbf{-0.517} &          -0.310 & \textbf{-0.615} &          -6.044 &  & CRPS              &          -1.519 &          -0.626 &          -0.872 &          -0.380 &                  -0.218 &          -0.537 &          -6.922 \\
			IS              &          -1.549 &          -0.438 &         -0.7030 & \textbf{-0.517} &          -0.310 &          -0.624 &          -6.054 &  & IS              &          -1.673 &          -0.592 &          -0.934 &          -0.381 &                  -0.203 &          -0.710 & \textbf{-5.493} \\ \hline\hline
		\end{tabular}			}
	\caption{{ Predictive accuracy of GVP using a Bayesian
		neural network for data generated from the dynamic non-linear regression
		model discussed in Section \protect\ref{sect:BNN}. Panels A to D document
		results for the four different versions of $z_t$, in which varying numbers
		of input variables are included in the predictive model. {The rows in each
			panel refer to the update method (U.method) used. The columns refer to the
			out-of-sample measure used to compute the average scores. The figures in
			bold are the largest average scores according to a given out-of-sample
			measure.}}}\label{tab:NNcovarites}
		\end{sidewaystable}

\clearpage

With reference to the results in Table~\ref{tab:NNcovarites} we can make two
observations. First, as anticipated, an increase in the information set
produces higher average scores out-of-sample, for all updates; with the
corresponding values increasing steadily and uniformly as one moves from the
results in Panel A (based on $z_{t}=y_{t-1}$) to those in Panel D (based on
the largest information, $z_{t}=\left( y_{t-1},x_{1,t},x_{2,t}\right)
^{\prime }$) set. Secondly however, despite the improved performance of all
versions of GVP as the information set is increased to better match that
used in the true DGP, strict coherence still prevails. That is, `focusing'
on the measure that matters in the construction of the GVP update, still
reaps benefits, despite the reduction in misspecification.

\section{Empirical Application: M4 Competition\label{emp}}

The Makridakis 4 (M4) forecasting competition was a forecasting event first
organised by the University of Nicosia and the New York University Tandon
School of Engineering in 2018. The competition sought submissions of point
and interval predictions at different time horizons, for a total of 100,000
times series of varying frequencies. The winner of the competition in a
particular category (i.e. point\ prediction or interval prediction) was the
submission that{\ achieved the best average out-of-sample predictive
	accuracy according to the measure of accuracy that defined that category,
	over all horizons and all series.\footnote{%
		Details of all aspects of the competition can be found via the following
		link:\newline
		https://www.m4.unic.ac.cy/wp-content/uploads/2018/03/M4-Competitors-Guide.pdf%
}}

We gauge the success of our method of\textbf{\ }\textit{distributional}
prediction in terms of the measure used to rank the interval forecasts in
the competition, namely the IS. We focus on one-step-ahead
prediction of each of the 4227 time series of daily frequency. Each of these
series is denoted by $\left\{ Y_{i,t}\right\} $ where $i=1,\dots ,4227$ and $%
t=1,\dots ,n_{i}$, and the task is to construct a predictive interval for $%
Y_{i,n_{i}+1}$ based on GVP. We adopt the mixture distribution in %
\eqref{Eq:misturepredclass} as the predictive model, being suitable as it is
to capture the stylized features of high frequency data. However, the model
is only appropriate for stationary time series and most of the daily time
series exhibit non-stationary patterns. To account for this, we model the
differenced series $\left\{ Z_{i,t}=\Delta ^{d_{i}}Y_{i,t}\right\} $, where $%
d_{i}$ indicates the differencing order. The integer $d_{i}$ is selected by
sequentially applying the KPSS test (\citealp{KPSS}) to $\Delta ^{j}y_{i,t}$
for $j=0,\dots ,d_{i}$, with $d_{i}$ being the first difference at which the
null hypothesis of no unit root is not rejected.\footnote{%
	To do this we employ the autoarima function in the forecast R package.} To
construct the predictive distribution of $Z_{i,n_{i}+1}$ we first obtain $%
M=5000$ draws $\{\theta _{i}^{(m)}\}_{m=1}^{M}$ from the Gibbs variational
posterior $q_{\tilde{\lambda}}\left( \theta _{i}\right) $ that is based 
on the IS.\footnote{%
	Once again, a value of $w=1$ is used in defining the Gibbs posterior and,
	hence, in producing the posterior approximation.} Draws $%
\{z_{i,n_{i}+1}^{(m)}\}_{m=1}^{M}$ are then obtained from the corresponding
predictive distributions $\{P_{\theta _{i}^{(m)}}^{(n_{i}+1)}\}_{m=1}^{M}$.
The draws of $Z_{i,n_{i}+1}$ are then transformed into draws of $%
Y_{i,n_{i}+1}$ and a predictive distribution for $Y_{i,n_{i}+1}$ produced
using kernel density estimation; assessment is performed in terms of the
accuracy of the prediction interval for $Y_{i,n_{i}+1}.$ 
To account for different units of measurement, the IS of each series
is then divided by the constant $\frac{1}{(n_i-1)}\sum_{t=2}^{n_i}|Y_{i,t}-Y_{i,t-1}|$.

Table~\ref{tab:m4msis} documents the predictive performance of our approach
relative to the competing methods \citep[see]{makridakis2020m4} for
details on these methods). The first column corresponds to the average IS.
In terms of this measure the winner is the method proposed by \cite%
{smyl2019hybrid}, while our method is ranked 11 out of 12. The ranking is\
six out of 12 in terms of the median IS, recorded in column two.
However, it is worth remembering that our method aims to achieve high
individual, and not aggregate (or average) predictive interval accuracy
across series. A more appropriate way of judging the effectiveness of our
approach is thus to count the number of series for which each method
performs best. This number is reported in column three of the table. In
terms of this measure, with a total number of 858 series, our approach is
only outperformed by the method proposed in \cite{doornik2019card}. In
contrast, although \citeauthor{smyl2019hybrid} is the best method in terms
of mean IS, it is only best at predicting 130 of the series in
the dataset. It is also important to mention that most of the competing
approaches have more complex assumed predictive classes than our mixture
model. Even the simpler predictive classes like the ARIMA and ETS models
have model selection steps that allow them to capture richer Markovian
processes than the mixture model which, by construction, is based on an
autoregressive process of order one.

In summary, \textit{despite} a single specification being defined for the
predictive class for all 4227 daily series, the process of driving
the Bayesian update via the IS rule has still yielded the \textit{%
	best}\textbf{\textit{\ }}predictive results in a very high number of cases,
with GVP beaten in this sense by only one other competitor. Whilst the
predictive model clearly matters, designing a bespoke updating mechanism to
produce the predictive distribution is shown to still reap substantial
benefits. 

\clearpage

\begin{table}[h!]
	\centering
	\begin{tabular}{lccc}
		\toprule\toprule & \multicolumn{3}{c}{\textbf{Out-of-sample results}} \\ 
		& \multicolumn{1}{l}{Mean IS} & \multicolumn{1}{l}{Median IS} & 
		\multicolumn{1}{l}{Series} \\ 
		\cmidrule{2-4} \textbf{Method} &  &  &  \\ 
		\cmidrule{1-1} &  &  &  \\ 
		\citeauthor{doornik2019card} & -9.36 & -4.58 & 2060 \\ 
		Mixture & -16.66 & -5.85 & 858 \\ 
		Trotta & -11.41 & -7.19 & 376 \\ 
		ARIMA & -10.07 & -5.67 & 278 \\ 
		\citeauthor{fiorucci2020groec} & -9.20 & -5.73 & 163 \\ 
		\citeauthor{smyl2019hybrid} & -8.73 & -6.65 & 139 \\ 
		Roubinchtein & -13.08 & -7.35 & 97 \\ 
		Ibrahim & -38.81 & -6.33 & 88 \\ 
		ETS & -9.13 & -5.71 & 67 \\ 
		\citeauthor{petropoulos2020simple} & -9.77 & -5.68 & 59 \\ 
		Montero-Manso, \textit{et al.} & -10.24 & -6.90 & 23 \\ 
		Segura-Heras, \textit{et al.} & -15.29 & -8.48 & 19 \\ 
		\bottomrule\bottomrule &  &  & 
	\end{tabular}%
	\caption{{ One-step-ahead predictive performance for
		the 4,227 daily series from the M4 competition. The columns `Mean IS' and
		Median IS' respectively record the mean and median values of average
		out-of-sample MSMS, across all 4,227 series, for each competiting
		method/team. The column `Series' reports the number of series for which the
		method/team produces the largest IS value. Some of the competing methods
		are individually described in \protect\cite{doornik2019card}, \protect\cite%
		{fiorucci2020groec}, \protect\cite{smyl2019hybrid}, \protect\cite%
		{petropoulos2020simple}, and \protect\cite{montero2019fforma}. For the
		remaining methods see \protect\cite{makridakis2020m4}. Our GVP method based
		on the mixture model is referred to as `Mixture'.}}\label{tab:m4msis}
\end{table}

\section{Discussion\label{disc}}

We have developed a new approach for conducting loss-based Bayesian
prediction in high-dimensional models. Based on a variational approximation
to the Gibbs posterior defined, in turn, by the predictive loss that is
germane to the problem at hand, the method is shown to produce predictions
that minimize that loss out-of-sample. `Loss' is characterized in this paper
by {positively-oriented proper scoring} rules designed to reward the
accuracy of predictive probability density functions for a continuous random
variable. Hence, loss minimization translates to maximization of an expected
score. However, in principle, any loss function in which predictive accuracy
plays a role could be used to define the Gibbs posterior. Critically, we
have proven theoretically, and illustrated numerically, that for a large
enough sample there is no loss incurred in predictive accuracy as a result
of approximating the Gibbs posterior.

{In comparison with the standard approach based on a likelihood-based
	Bayesian posterior, }our Gibbs variational predictive approach is ultimately
aimed at generating accurate predictions in the realistic empirical setting
where the predictive model and, hence, the likelihood function is
misspecified. Gibbs variational prediction enables the investigator to break
free from the shackles of likelihood-based prediction, {and to drive }%
predictive outcomes according to the form of predictive accuracy that
matters for the problem at hand; and all with theoretical validity
guaranteed.

We have focused in the paper on particular examples where the model used to
construct the Gibbs variational predictive is observation-driven. Extensions
to parameter-driven models (i.e., hidden Markov models, or state space
models) require different approaches with regard to both the implementation
of variational Bayes, and in establishing the asymptotic properties of the
resulting posteriors and predictives. This is currently the subject of
on-going work by the authors, \cite{frazier2021note}, and we reference \cite{tran2017variational} and 
\cite{quiroz2018gaussian} for additional discussion of this problem.

This paper develops the theory of Gibbs variational prediction for classes
of models that are general and flexible enough to accommodate a wide range
of data generating processes, and which, under the chosen loss function, are
smooth enough to permit a quadratic expansion. This latter condition
restricts the classes of models, and loss functions, under which our results
are applicable. For example, the discrete class of models studied in \cite%
{douc2013ergodicity} may not be smooth enough to deduce the validity of such
an expansion. 

While our approach to prediction focuses on accuracy in a given forecasting
rule, which is related to the notion of sharpness in the probabilistic
forecasting literature (see, \citealp{gneiting2007probabilistic}), it does
not pay attention to the calibration of such predictive densities. That is,
there is no guarantee that the resulting predictive densities have valid
frequentist coverage properties. If calibration is a desired property, it is
possible to augment our approach to prediction with the recently proposed
approach by \cite{wu2021calibrating}, which calibrates generalized
predictives. The amalgamation of these two procedures should produce accurate
predictions that are well-calibrated in large samples. We leave an
exploration of this possibility to future research.


\acks{We would like to thank various participants at: the `ABC in Svalbard' Workshop,
	April 2021, the International Symposium of Forecasting, the Alan Turing
	Institute, and RIKEN reading group, June 2021, the European Symposium of
	Bayesian Econometrics, September 2021, and the Vienna University of
	Business, December 2021, for helpful comments on earlier drafts of the
	paper. This research has been supported by Australian Research Council (ARC)
	Discovery Grants DP170100729 and DP200101414. Frazier was also supported by
	ARC Early Career Researcher Award DE200101070.}



\appendix 

\section{Theoretical Results}

\label{app:theory} In this section, we state several intermediate results
and prove Theorems \ref{thm:merge1} and \ref{thm:merge} stated in Section %
\ref{theory} of the main paper.

We maintain the following notation throughout the remainder of the paper. We
let $d:\Theta \times \Theta \rightarrow \mathbb{R}_{x\geq 0}$ denote a
generic metric on $\Theta\subseteq\mathbb{R}^{d_n}$, and we\textbf{\ }take $%
\Vert \cdot \Vert $ and $\langle \cdot ,\cdot \rangle $ to be the Euclidean
norm and inner product for vectors. In the following section, we let $C$ denote a generic constant
independent of $n$ that can vary from line to line. For sequences $%
x_{n},y_{n}$ and some $C$ we write $x_{n}\lesssim y_{n}$ when $x_{n}\leq
Cy_{n}$. If $x_{n}\lesssim y_{n}$ and $y_{n}\lesssim x_{n}$ we write $%
x_{n}\asymp y_{n}$. For some positive sequence $\delta _{n}$, the notation $%
o_{p}(\delta _{n})$ and $O_{p}(\delta _{n})$ have their usual connotation.
Unless otherwise stated, all limits are taken as $n\rightarrow \infty $, so
that when no confusion will result we use $\lim_{n}$ to denote $%
\lim_{n\rightarrow \infty }$. The random variable $Y_{n}$ takes values in $%
\mathcal{Y}$ for all $n\geq 1$, and for a given $n$, the vector of
observations $y_{1:n}$ are generated from the true probability measure $%
P_{0} $, not necessarily in $\mathcal{P}^{(n)}$, and whose dependence on $n$
we suppress to avoid confusion with the predictive model $P_{\theta }^{(n)}$.

\subsection{Assumptions and Discussion}

For some positive sequence $\epsilon _{n}\rightarrow 0$, define $\Theta
(\epsilon _{n}):=\{\theta \in \Theta :d(\theta ,\theta _{\star })\leq
\epsilon _{n}\}$ as an $n$-dependent neighbourhood of $\theta _{\star }$ and
let $\Theta ^{c}(\epsilon _{n})$ denote its complement.

\begin{assumption}
	\label{ass:est} (i) There exists a non-random function $\mathcal{S}(\theta )$
	such that $\sup_{\theta \in \Theta }|S_{n}(\theta )/n-\mathcal{S} (\theta )|
	\rightarrow 0$ in probability. (ii) For any $\epsilon \geq \epsilon _{n}$,
	there exists some $C>0$ such that $\sup_{\Theta^{c} (\epsilon )}\left\{ 
	\mathcal{S}(\theta )-\mathcal{S}(\theta _{\star })\right\} \leq -C\epsilon
	^{2}$.
\end{assumption}

\begin{rem}
	Assumption \ref{ass:est} places regularity conditions on the sample and
	limit counterpart of the expected scoring rules, $\mathcal{S}(\theta)$, and is used to deduce the
	asymptotic behavior of the Gibbs posterior. The first part of the assumption
	is a standard uniform law of large numbers, and the second part amounts to
	an identification condition for $\theta _{\star }$.
\end{rem}

We note here that the temporal dependence of the observations, and the use
of potentially heavy-tailed data means that existing results on the
concentration of Gibbs posteriors, and/or their variational approximation,
of which we are aware may not be applicable in our setting; see e.g., \cite%
{zhang2020convergence}, \cite{alquier2020concentration} and \cite%
{yang2020alpha}.

Instead, we use smoothness of $S_{n}(\theta )$ (and the underlying model $%
P_{\theta }^{(n)}$) to deduce a posterior concentration result.

\begin{assumption}
	\label{ass:one}There exists a sequence of $d_n\times d_n$-dimensional matrices $%
	\Sigma_n$, and a $d_n\times1$-random vector $\Delta_n$ such:
	
	\begin{enumerate}
		\item[(i)] $S_{n}({\theta })-S_{n}({\theta }_{\star })=\langle \sqrt{n}
		\left( {\theta }-{\ \theta _{\star }}\right) ,{\Delta }_{n}/\sqrt{n}\rangle
		- \frac{1}{2}\Vert \Sigma _{n}^{1/2}\sqrt{n}\left( {\theta }-{\theta }%
		_{\star }\right) \Vert ^{2}+R_{n}({\theta }).$
		
		\item[(ii)] For any $\epsilon >\epsilon _{n}$, any $M>0$, with $M/\sqrt{n}
		\rightarrow 0$, $\lim \sup_{{}}P_{0}^{}\left[ \sup_{d(\theta ,\theta _{\star
			})\leq M/\sqrt{n}}\frac{\left\vert R_{n}(\theta )\right\vert}{%
			1+n\|\theta-\theta_\star\|^2} >\epsilon \right] =0.$
		
		\item[(iii)] $\left\Vert \Sigma _{n}^{-1/2}\Delta _{n}\right\Vert =O_{p}(1)$.
	\end{enumerate}
\end{assumption}

We require the following a tail control condition on the prior.

\begin{assumption}
	\label{ass:post} For any $\epsilon>\epsilon_n$, $\log\left\{\Pi\left[
	\Theta(\epsilon)\right]\right\}\gtrsim -n\epsilon^2$.
\end{assumption}

\begin{rem}
	The above assumptions ultimately preclude cases where $\Theta $ can be
	partitioned into global parameters that drive the model, which are fixed for
	all $n$, and local parameters that represent time-dependent latent variables
	in the model and grow in a one-for-one manner with the sample size. As such,
	these assumptions are not designed to be applicable to variational inference
	in classes such as hidden Markov models.
\end{rem}

To validate the usefulness and broad applicability of Assumptions \ref%
{ass:est}-\ref{ass:one}, we prove that these assumptions are satisfied for
the GARCH(1,1) model used for the illustrative exercise in Section \ref{sec:toy} in the case of
the logarithmic scoring rule, $S(y,f)=\log f(y)$, under weak
moment assumptions on the observed data.

\begin{lem}
	\label{lem:garch} Recall the GARCH(1,1) model presented in Section \ref%
	{sec:toy}, and assume, WLOG, that $\mathbb{E}[Y_t]=0$. Define the rescaled
	variable $Z_t=Y_t/\sigma_{t\star}$, where $\sigma_{t\star}$ denotes the
	GARCH process evaluated at $\theta=\theta_\star$. Assume that $Z_t$
	satisfies the following conditions:
	
	\begin{enumerate}
		\item $Z_t$ is strictly stationary and ergodic.
		
		\item $Z_t^2$ is a random variable.
		
		\item For some $\delta>0$, there exists an $C<\infty$ such that $\mathbb{E}%
		[Z_t^{4}|\mathcal{F}_{t-1}]\le C<\infty$.
		
		\item The parameter space $\Theta$ is compact, with the elements $%
		\theta_2,\theta_3,\theta_4$ elements bounded away from zero and the elements 
		$\theta_3,\theta_4$ bounded above by one, and where $\theta_3+\theta_4<1$.
		The pseudo-true value $\theta^\star$ is in the interior of $\Theta$.
	\end{enumerate}
	
	If the above are satisfied, then Assumptions \eqref{ass:est} and %
	\eqref{ass:one} are satisfied for $S_n(\theta)=\sum_{t=2}^{n}\log p_\theta
	(y_t|\mathcal{F}_{t-1})$.
\end{lem}

\subsection{Preliminary Result: Posterior Concentration}

Before proving the results stated in Section \ref{theory}, we first give two results regarding the
posterior concentration of the exact Gibbs posterior and its variational
approximation.

\begin{lem}
	\label{lemma:conv} Under Assumptions \ref{ass:est}-\ref{ass:post}, for $%
	\epsilon_{n}\rightarrow0$, $n\epsilon_n^2\rightarrow\infty$, and any $%
	M_n\rightarrow\infty$, for $w_n\in(\underline{w},\overline{w})$ with
	probability one, where $0<\underline{w}<\overline{w}<\infty$, 
	\begin{equation*}
	\Pi_{w_n}\left(d(\theta,\theta_\star)>M_n\epsilon_n|y_{1:n}\right)\lesssim
	\exp\left(-CM_n^2\epsilon_n^2n\right),
	\end{equation*}
	with probability converging to one.
\end{lem}

\begin{rem}
	The proof of Lemma \ref{lemma:conv}, requires controlling the behavior of $%
	\sup_{d(\theta,\theta_\star)\ge\epsilon}n^{-1}\left\{S_n(\theta)-S_n(\theta_
	\star)\right\} $ for any $\epsilon>0$. When the parameter space is compact,
	this control can be guaranteed by controlling the bracketing entropy (see,
	e.g., \citealp{vdv1996}) of $\Theta$. If the parameter space is not compact,
	this control can be achieved by considering more stringent conditions on the
	prior than Assumption \ref{ass:post}. In particular, results where the
	parameter space is not compact can be obtained by modifying Theorem 4 in 
	\cite{shen2001rates} for our setting. {That being said, we note that in the
		case of the toy GARCH(1,1) model in Section \ref{sec:toy}, the parameter
		space is compact and this condition is satisfied under the conditions given
		in Lemma \ref{lem:garch}.}
\end{rem}

To transfer the posterior concentration of the Gibbs posterior to its
variational approximation, we restrict the class $\mathcal{Q}$ used to
produce the variational approximation. Following \cite{zhang2020convergence}%
, we define $\kappa _{n}$ to be the approximation error rate for the
variational family: 
\begin{equation*}
\kappa _{n}^{2}=\frac{1}{n}\E_{P_0}\inf_{Q\in \mathcal{Q}}\text{D}_{ }\left\{
Q\Vert \Pi _{w}(\cdot |y_{1:n})\right\} .
\end{equation*}

\begin{lem}
	\label{thm:one} Under the Assumptions of Lemma \ref{lemma:conv}, for any
	sequence $M_n\rightarrow\infty$, 
	\begin{equation*}
	\widehat{Q}\left\{d(\theta,\theta_\star)\geq M_n
	n\left(\epsilon^2_n+\kappa^2_n\right)|y_{1:n}\right\}\rightarrow 0
	\end{equation*}
	in probability.
\end{lem}

\begin{rem}
	While several authors, such as \cite{Alquier2016}, \cite%
	{alquier2020concentration}, and \cite{yang2020alpha}, have demonstrated a
	result similar to Lemma \ref{thm:one}, existing results generally hinge on
	the satisfaction of {a sub-Gaussian} concentration inequality for the
	resulting risk function, i.e., $S_n(\theta)$, used to produce the Gibbs
	posterior, such as a Hoeffding or Bernstein-type inequality. Given that our
	goal is prediction in possibly non-Markovian, non-stationary, and/or complex
	models, and with a loss function based specifically on a general scoring
	rule, it is not clear that such results {remain applicable}. Therefore, we
	deviate from the earlier referenced approaches and {instead draw inspiration
		from \cite{miller2021asymptotic}, and use an approach based on a quadratic
		expansion of the loss function. The latter conditions can be shown to be
		valid in a wide variety of models used to produce predictions in commonly
		encountered time series settings, including models with non-Markovian
		features}. 
	%
	Indeed, as Lemma \ref{lem:garch} demonstrates, Assumptions \ref{ass:est} and %
	\ref{ass:one} are satisfied in cases where the scoring rule only exhibits a
	polynomial tail, e.g., in the case of the GARCH(1,1) model, and with data that is non-markovian.
\end{rem}

\begin{rem}
	Lemma \ref{thm:one} gives a similar result to Theorem 2.1 in \cite%
	{zhang2020convergence} but for the Gibbs variational\textbf{\ }posterior and
	demonstrates that the latter concentrates onto $\theta _{\star }$, with the
	rate of concentration given by the slower of $\epsilon _{n}$ and $\kappa
	_{n} $. We note that, in cases where the dimension of $\Theta $ grows with
	the sample size, the rate $\epsilon _{n}$ is ultimately affected by the rate
	of growth of $\Theta $ and care must be taken to account for this
	dependence. 
\end{rem}

\begin{rem}
	Verification of Lemma \ref{thm:one} in any practical example requires
	choosing the variational family $\mathcal{Q}$. In the case of
	observation-driven time series models, where the predictive model $P_{\theta
	}^{(n)}$ can be constructed analytically, and no explicit treatment of
	`local' parameters is needed, the variational family can be taken to be a
	sufficiently rich parametric class. For example, consider the case where the
	dimension of $\Theta $ is fixed and compact, and assume that $P_{\theta }^{(n)}$ and $S_n(\theta)$ satisfy the Assumptions of Lemma \ref{lemma:conv}. Then, we can consider
	as our variational family the mean-field class: 
	\begin{equation*}
	\mathcal{Q}_{MF}:=\left\{ Q:q(\theta )=\prod_{i=1}^{d_{\theta }}q_{i}(\theta
	_{i})\right\} ,
	\end{equation*}
	or the Gaussian family 
	\begin{equation*}
	\mathcal{Q}_{G}:=\{\lambda =(\mu ^{\prime },\text{vech}({\Sigma })^{\prime
	})^{\prime }:q(\theta )\propto \mathcal{N}(\theta ;\mu ,\Sigma ),\},
	\end{equation*}
	where $\mu \in \mathcal{R}^{d}$, and $\Sigma $ is a $d\times d$
	positive-definite matrix. In either case, the approximation results of \cite%
	{zhang2020convergence} and \cite{yang2020alpha} for these variational
	families can be used to obtain the rate $\kappa _{n}$. In particular, when $%
	d_{ }$ is fixed these results demonstrate that $\kappa _{n}\lesssim
	\epsilon _{n}$, and the Gibbs variational posterior converges at the same
	rate as the Gibbs posterior.\footnote{%
		The rate of concentration obtained by applying Lemma \ref{lemma:conv} in
		such an example will be $\epsilon _{n}=\log (n)/\sqrt{n}$, which is slightly
		slower than the optimal parametric rate $\epsilon _{n}=1/\sqrt{n}$. The
		parametric rate can be achieved by sufficiently modifying the prior
		condition in Assumption \ref{ass:post} to cater for the parametric nature of
		the model. However, given that this is not germane to the main point off the
		paper, we do not analyze such situations here.}
\end{rem}

As seen from Lemma \ref{thm:one}, the variational posterior places
increasing mass on the value $\theta _{\star }$ that maximizes the limit of%
\textbf{\ }the expected scoring rule, $\mathcal{S}(\theta)$. Lemma \ref{thm:one} does not rely on
the existence of exponential testing sequences as in much of the Bayesian
nonparametrics literature (see, e.g., the treatment in %
\citealp{ghosal2017fundamentals}). The need to consider an alternative
approach is due to the loss-based, i.e., non-likelihood-based, nature of the
Gibbs posterior. Instead, the arguments used to demonstrate posterior
concentration rely on controlling the behavior of a suitable quadratic
approximation to the criterion function in a neighborhood of $\theta _{\star
}$.

The result of Lemma \ref{thm:one} allows us to demonstrate that the
predictions made using the GVP, $P_{Q}^{(n)}$, are just as accurate as those
obtained by an agent that uses the `optimal [frequentist] predictive' $P_{\star }^{(n)}$ defined in equation \eqref{eq:limit_loss};
this is the key message of \ref{thm:merge}, which demonstrates that the GVP
yields predictions that are just as accurate as the exact Gibbs predictive
(in large samples).

\subsection{Proofs of Lemmas}

\begin{proof}{\bf of  Lemma \ref{lem:garch}.} We break the proof up by showing the results for Assumption \ref{ass:est} and \ref{ass:one} separately. 
	
	\bigskip

	\noindent\textbf{Assumption \ref{ass:est}.}	 We remark that Assumptions 1-4 in Lemma \ref{lem:garch} are sufficient for conditions (A.1) and (A.2) of \cite{lee1994asymptotic}. To establish Assumption \ref{ass:est}(i) we note that Lemma 7 in \cite{lee1994asymptotic}(2) implies that $S_n(\theta)\rightarrow_p\mathcal{S}(\theta)$ for each $\theta\in\Theta$. Furthermore, since the expected value of $\partial \log p_\theta (y_t|\mathcal{F}_{t-1})/\partial\theta$ is bounded for each $t$, for each $\theta\in\Theta$, by Lemma 8(2) in \cite{lee1994asymptotic}, it follows that $S_n(\theta)\rightarrow_p\mathcal{S}(\theta)$ uniformly over $\Theta$. Assumption \ref{ass:est}(ii) can then be established by using the fact that $\mathcal{S}(\theta)$ is twice-continuously differentiable for $\theta$ near $\theta_\star$ with negative-definite second derivative matrix $\partial^2\mathcal{S}(\theta_\star)/\partial\theta\partial\theta'$, by Lemma 11(3) of \cite{lee1994asymptotic}, which yields
	$$
	\mathcal{S}(\theta)-\mathcal{S}(\theta^\star)\le -C\|(\theta-\theta_\star)\|^2,
	$$for some $C>0$. Hence, Assumption \ref{ass:est}(ii) follows. 
	
	\bigskip

	\noindent\textbf{Assumption \ref{ass:one}.}	
	The result again follows by using the results of \cite{lee1994asymptotic}. In particular, in this case $S_n(\theta)=\sum_{t=1}^{n}\log p_\theta(y_{t+1}|\mathcal{F}_t)$, is twice continuously differentiable so that a direct quadratic approximation is valid at $\theta=\theta^\star$. Then, taking $\Delta_n=\partial S_n(\theta_\star)/\partial\theta$ and $\Sigma_n=-\partial^2\mathcal{S}(\theta_\star)/\partial\theta\partial\theta'$ yields the expansion, where the remainder term is given by
	\begin{flalign*}
	R_n(\theta)&=n(\theta-\theta_\star)'[\partial^2\mathcal{S}(\theta_\star)/\partial\theta\partial\theta'-\partial^2\mathcal{S}(\bar\theta)/\partial\theta\partial\theta'](\theta-\theta_\star)\\&-n(\theta-\theta_\star)'[n^{-1}\partial^2S_n(\bar\theta)/\partial\theta\partial\theta'-\partial^2\mathcal{S}(\bar\theta)/\partial\theta\partial\theta'](\theta-\theta_\star),
	\end{flalign*}and $\bar\theta$ an intermediate value such that $\|\theta-\bar\theta\|\le\|\theta-\theta_\star\|$. 
	
	To verify (ii), we note that $\partial^2\mathcal{S}(\bar\theta)/\partial\theta\partial\theta'$ exists and is continuous by Lemma 11(3) of \cite{lee1994asymptotic}, so that the first term is $o(1)$ on $d(\theta,\theta_\star)\le M/\sqrt{n}$. Further, Lemma 11(3) of \cite{lee1994asymptotic} demonstrates that 
	$$
	\sup_{\theta\in\Theta}\|[n^{-1}\partial^2S_n(\bar\theta)/\partial\theta\partial\theta'-\partial^2\mathcal{S}(\bar\theta)/\partial\theta\partial\theta']\|=o_p(1),
	$$so that the second term is also $o(1)$ on $d(\theta,\theta_\star)\le M/\sqrt{n}$.
	
	Lastly, (iii), is verified since $\Sigma_n=-\partial^2\mathcal{S}(\theta_\star)/\partial\theta\partial\theta'$, and since, by Lemma 9(2) of \cite{lee1994asymptotic}, $\Delta_n/\sqrt{n}\Rightarrow N(0,\E\left[\partial \log p_\theta (y_t|\mathcal{F}_{t-1})/\partial\theta\partial \log p_\theta (y_t|\mathcal{F}_{t-1})/\partial\theta'\right])$.
	
\end{proof}

\begin{proof}{\bf of  Lemma \ref{lemma:conv}.}
	The posterior density is 
	$
	\Pi_{w_n}\left(\theta|y_{1:n}\right)\propto \dt\Pi(\theta)\exp\left[w_n\{{S}_n(\theta)\}\right],
	$where we recall that, by hypothesis, $w_n\in(\underline{w},\overline{w})\subset (0,\infty)$ with probability one. Define the quasi-likelihood ratio 
	$$
	Z_n(\theta):= \exp\left[w_n\left\{{S}_n\left(\theta\right)-{S}_n(\theta_\star)-\frac{1}{2}\Delta_{n}'\Sigma^{-1}_n\Delta_{n}\right\}\right],
	$$where, by Assumption \ref{ass:one} (iii), $\Sigma_n^{-1/2}\Delta_n=O_p(1)$. For $M>0$, the posterior probability over $A_n:=\{\theta:d(\theta,\theta_\star)>M\epsilon_n\}$ is
	\begin{flalign*}
	\Pi_{w_n}(A_n|y_{1:n})=\frac{\int_{A_n}\dt\Pi\left(\theta\right) Z_{n}(\theta)}{\int_{\Theta} \dt\Pi\left(\theta\right) Z_{n}(\theta)}=\frac{N_n(A_n)}{D_n},
	\end{flalign*}where 
	$$
	{N_n(A_n)}=\int_{A_n}\dt\Pi\left(\theta\right) Z_{n}(\theta), \quad {D_n}={\int_{\Theta} \dt\Pi\left(\theta\right) Z_{n}(\theta)}.
	$$
	The result now follows by lower bounding $D_n$ and Upper bounding $N_n(A_n)$. 
	
	\medskip

	\noindent\textbf{Part 1: $D_n$ Term.}
	Define $T_n:=\theta_\star+\Sigma_n^{-1}\Delta_{n}$, and $%
	G_n:=\{\theta\in\Theta:\frac{1}{2}(\theta-T_n)^{\prime }\left[\Sigma_n^{}/n %
	\right](\theta-T_n)\leq t_n\}$. We show that, for $t_n\rightarrow0$ , and $t_n\le
	\epsilon_n^2$, wpc1 
	\begin{equation*}
	D_n=\int_{\Theta}\dt\Pi(\theta)Z_n(\theta)>\frac{1}{2}\Pi(G_n)e^{-2\overline{
			w}nt_n}.
	\end{equation*}
	
		\bigskip

	\begin{proof}{\textbf{of Part 1:}} To lower bound $D_n$ we follow an approach similar approach to Lemma 1 in \cite{shen2001rates} (see, also, \citealp{syring2020gibbs}).	Over $G_n$, 
		use Assumption \ref{ass:one}(i) to rewrite $\log\{Z_n(\theta)\}$ as 
		\begin{flalign}
		\log\{Z_n(\theta)\}&=\log\left[\exp\left\{w_n\left[S_n(\theta)-S_n(\theta_\star)-\frac{1}{2}\Delta_{n}'\Sigma^{-1}_n\Delta_{n}\right]\right\}\right]\nonumber\\&=-\frac{w_nn}{2}(\theta-T_n)'\left[\Sigma_n/n\right](\theta-T_n)-R_n(\theta).\label{eq:app3}
		\end{flalign}
		Define the following sets: $\mathcal{C}_n:=\{(\theta,y_{1:n}):|R_n(\theta)|\ge n t_n\}$,  $\mathcal{C}_n(\theta):=\{y_{1:n}:(\theta,y_{1:n})\in\mathcal{C}_n\}$, and $\mathcal{C}_n(y_{1:n}):=\{\theta:(\theta,y_{1:n})\in\mathcal{C}_n\}$. On the set $G_n\cap\mathcal{C}^{c}_n(y_{1:n})$, $Z_n(\theta)$ is bounded in probability, and we can bound $D_n$ as follows:
		\begin{flalign}
		D_n&\geq \int_{G_n\cap\mathcal{C}^{c}_n(y_{1:n})}\exp\left\{-\frac{w_nn}{2}(\theta-T_n)'[\Sigma_n/n](\theta-T_n)-R_n(\theta)\right\}\dt\Pi(\theta)\nonumber\\&\geq \exp\left(-2\overline{\omega} nt_n\right)\Pi\left\{G_n\cap\mathcal{C}^{c}_n(y_{1:n}) \right\}\nonumber\\&\geq\left[\Pi(G_n)-\Pi\left\{G_n\cap \mathcal{C}_n(y_{1:n})\right\}\right]\exp\left(-2\overline{\omega} nt_n\right).
		\label{eq:bound}
		\end{flalign}From the bound in \eqref{eq:bound}, the remainder follows almost identically to Lemma 1 in \cite{shen2001rates}. In particular, 
		\begin{flalign*}
		\E_{P_{0}}\left\{\Pi\left[G_n\cap\mathcal{C}_n(y_{1:n})^{}\right]\right\}&=\int_{\mathcal{Y}} \int_{\Theta} \I\{ G_n\cap \mathcal{C}_n(y_{1:n})^{} \}\dt\Pi(\theta)\dt P_{0}^{}(y_{1:n})\\&=\int_{\mathcal{Y}} \int_{\Theta} \I\{G_n\}\I\{\mathcal{C}_n(y_{1:n})^{}\}\dt\Pi(\theta)\dt P_{0}^{}(y_{1:n})\\&\leq \frac{1}{nt_n}\Pi(G_n),
		\end{flalign*}
		where the last line follows from Markov's inequality and the definition of $\mathcal{C}_n(y_{1:n})$. 	Lastly, consider the probability of the set 
		\begin{flalign*}
		P_{0}^{}\left\{D_n\leq \frac{1}{2}\Pi(G_n)e^{-2\overline{w}nt_n}\right\}&\leq P_{0}^{}\left(e^{-2\overline{w}nt_n}\left[\Pi(G_n)-\Pi\left\{G_n\cap \mathcal{C}_n(y_{1:n})^{}\right\}\right]\leq \frac{1}{2}\Pi(G_n)e^{-2\overline{w}nt_n}\right)\\&=P_{0}^{}\left[\Pi\{G_n\cap\mathcal{C}_n(y_{1:n})\}\ge \frac{1}{2}\Pi(G_n)\right]\\&\leq 2 P_{0}^{}\left\{G_n\cap \mathcal{C}_n(y_{1:n})\right\}/\Pi(G_n)\\&\leq \frac{2}{n t_n}.  
		\end{flalign*}Hence, 
		$
		P_{0}^{}\left\{D_n\geq \frac{1}{2}\Pi(G_n)e^{-2\overline{w}n t_n}\right\}\geq 1-2(nt_n)^{-1}$ and for $nt_n\rightarrow\infty$, we have that 
		\begin{flalign}\label{eq:dbnound}
		D_n\geq \frac{1}{2}\Pi(G_n)e^{-2\overline{w}n t_n} 
		\end{flalign}
		except on sets of $P_{0}^{}$-probability converging to zero. 
	\end{proof}

	Then, for $G_n$ as in that result, for a sequence $t_n\rightarrow0$ with $t_n\le \epsilon_n^2$, by Assumption \ref{ass:post}  (wpc1)
	\begin{flalign}
	D_n\ge \frac{1}{2}\Pi(G_n)e^{-2nt_nw_n}\gtrsim e^{-C_1nt_n}e^{-2nt_nw_n}\gtrsim e^{-(C_1+2\overline{w})nt_n}\label{eq:lower}
	\end{flalign}

	\bigskip

	\noindent\textbf{Part 2: $N_n(A_n)$ Term.} We show that $P_0^{(n)}[N_n(A_n)>e^{-M\overline{w}n\epsilon^2_n}]\le e^{-C_2 M\overline{w}nt^2_{n}}$ for $n$ large enough. 
	
		\bigskip

	\begin{proof}{\textbf{of Part 2:}} 	
		To bound $N_n(A_n)=\int_{A_n}\dt\Pi(\theta)Z_n(\theta)$ from above, use Assumption \ref{ass:est} to deduce that, for any positive $\epsilon_n$, 
		\begin{flalign}
		\sup_{d(\theta,\theta_\star)\ge\epsilon_n}n^{-1}\left\{S_n(\theta)-S_n(\theta_\star)\right\}\leq& 2\sup_{\theta\in\Theta}\{S_n(\theta)/n-\mathcal{S}(\theta)\}+\sup_{d(\theta,\theta_\star)\ge \epsilon}\{\mathcal{S}(\theta)-\mathcal{S}(\theta_\star)\}\nonumber\\&\leq o_p(1)-C_2\epsilon^2_n.\label{eq:numerator}
		\end{flalign}
		For some $M>0$, we decompose $N_n(A_n)$: $$N_{n}(A_{n})=N_{n}(A_{n}\cap\Theta_{n}(M))+N_{n}(A_{n}\cap\Theta_{n}^{c}(M))=N_{n}^{(1)}+N_{n}^{(2)},$$
		where $\Theta_{n}(M)$ is a compact set of $\theta$ by construction and $\Theta_{n}^{c}(M)$ is
		its complement. 
		
		Then, $N_{n}^{(1)}\le e^{-C_2 M\overline{w}n\epsilon^2_{n}}$ from \eqref{eq:numerator} due to the compactness of $\Theta_n(M)$. For $N_n^{(2)}$,
		for a sequence $t_n\rightarrow0$ with $t_n\le \epsilon_n^2$, 
		\begin{flalign}
		&P_{0}(N_{n}^{(2)}>e^{-M \overline{w} n t_{n}/2})\nonumber\\ 
		\le &  e^{M \overline{w} n t_{n}/2} \int_{\mathcal{Y}} \int_{A_{n}\cap\Theta_{n}^{c}(M)}\exp\left\{-\frac{w_nn}{2}(\theta-T_n)'[\Sigma_n/n](\theta-T_n)-R_n(\theta)\right\}\dt\Pi(\theta)\dt P_0(y_{1:n})\nonumber\\
		\leq & e^{M \overline{\omega} nt_n/2}\Pi\left\{A_{n}\cap\Theta_{n}^{c}(M) \right\} \leq e^{-MC_2 \overline{w} nt_{n}}\label{eq:upper}
		\end{flalign}
		The first inequality comes from Markov inequality and the definition of $Z_n(\theta)$ and the second and the third inequalities are due to Assumptions \ref{ass:one} and \ref{ass:post} respectively.
	\end{proof}

	Using \eqref{eq:lower}, we have the posterior bound	
	\begin{flalign*}
	\Pi_{w_n}(A_n|y_{1:n})\lesssim N_n(A_n)e^{(C_1+2\overline{w})nt_n}
	\end{flalign*}Combining $N_{n}^{(1)}\le e^{-C_2 M\overline{w}n\epsilon^2_{n}}$, from \eqref{eq:numerator}, and equation \eqref{eq:upper}, we can deduce that  (wpc1)
	\begin{flalign*}
	\Pi_{w_n}(A_n|y_{1:n})\lesssim N_n(A_n)e^{(C_1+2\overline{w})nt_n}\lesssim {e^{-\left\{MC_2\overline{w}-C_1-2\overline{w}\right\}n\epsilon^2_n}},
	\end{flalign*}since $t_n\le \epsilon_n^2$. The right-hand-side vanishes for $M$ large enough, and the result follows. 
	
\end{proof}

\begin{proof}{\bf of  of Lemma \ref{thm:one}.}
	The result begins with a similar approach to Corollary 2.1 of \cite{zhang2020convergence}, but requires particular deviations given the non-likelihood-based version of our problem. We also deviate from \cite{zhang2020convergence} and use $\E_{P_0}[f]$ to denote expectations of $f(\cdot)$ taken under $P_0$, rather than the notation $P_0[f]$. 
	
	First, we note that Lemma B.2 in the supplement to \cite{zhang2020convergence} (reproduced for convenience as Lemma \ref{lem:general1} in Appendix \ref{app:replems}) can be directly applied in this setting: for any $a>0$ and $n\ge 1$, given observations $y_{1:n}$, 
	$$
	a \widehat{Q}\left\{d(\theta,\theta_\star)\right\} \leq \text{D}\left[\widehat{Q} \| \Pi_{w_n}\left(\cdot | y_{1:n}\right)\right]+\log \Pi_{w_n}\left(\exp \left\{ad(\theta,\theta_\star)\right\} | y_{1:n}\right).
	$$Similarly, for any $Q\in\mathcal{Q}$, $\text{D}\left[{Q}\|\Pi_{w_n}(\cdot|y_{1:n})\right]+o_p(1)\geq\text{D}\left[\widehat{Q}\|\Pi_{w_n}(\cdot|y_{1:n})\right]$ by construction, so that
	\begin{equation}
	a \widehat{Q}\left\{d(\theta,\theta_\star)\right\} \leq \inf_{Q \in \mathcal{Q}} \text{D}\left[{Q} \| \Pi_{w_n}\left(\cdot | y_{1:n}\right)\right]+\log \Pi_{w_n}\left(\exp \left\{ad(\theta,\theta_\star)\right\} | y_{1:n}\right).\label{eq:app1}
	\end{equation}
	Taking expectations on both sides of \eqref{eq:app1}, and re-arranging terms, yields
	\begin{flalign}
	\E_{P_0}^{} \widehat{Q}\left\{d(\theta,\theta_\star)\right\} &\leq \frac{1}{a}\E_{P_0}^{}\left\{\inf_{Q \in \mathcal{Q}} \text{D}\left[{Q} \| \Pi_{w_n}\left(\cdot | y_{1:n}\right)\right]+\log \Pi_{w_n}\left(\exp \left\{ad(\theta,\theta_\star)\right\} | y_{1:n}\right)\right\}\nonumber\\&\leq \frac{1}{a}n\kappa^2_n+\frac{1}{a}\log\left[\E_{P_0}^{}\Pi_{w_n}\left(\exp \left\{ad(\theta,\theta_\star)\right\} | y_{1:n}\right)\right],\label{eq:app2}
	\end{flalign}where the second inequality follows from the definition of $\kappa_n^2$, and Jensen's inequality. 
	
	The second term in equation \eqref{eq:app2} is bounded by applying Lemma \ref{lemma:conv}. In particular,  for all $\alpha \geq \alpha_0>0$ and any $0<a\leq \frac{1}{2}c_1$, $c_1>0$, by Lemma \ref{lemma:conv} (wpc1),
	$$
	\Pi_{w_n}\left(\exp \left\{ad(\theta,\theta_\star)\right\}>\alpha_0 | y_{1:n}\right) \lesssim \exp \left(-a c_{1} \alpha\right).
	$$	
	Then, appealing to Lemma B.4 in the supplement to \cite{zhang2020convergence} (reproduced for convenience as Lemma \ref{lem:sub-exp} in Appendix \ref{app:replems}) we obtain 
	$$
	\E_{P_0}^{} \Pi_{w_n}\left(\exp \left\{a d(\theta,\theta_\star)\right\} | y_{1:n}\right) \lesssim \exp \left(a c_{1} \alpha_0\right),
	$$for all $a\leq \min\{c_1,1\}$. Taking $a=\min\{c_1,1\}$ and $\alpha_0=n\epsilon_n^2$, and applying the above in equation \eqref{eq:app2}, yields, for some $M>0$, 	 
	\begin{flalign*} \E_{P_0}^{} \widehat{Q}\left\{d(\theta,\theta_\star)\right\} & \leq \frac{n \kappa_{n}^{2}+\log \left(c_1+e^{a c_{1} n \epsilon_{n}^{2}}\right)}{a} \lesssim {n \kappa_{n}^{2}}+ n \epsilon_{n}^{2}+o(1) \lesssim n\left(\kappa_{n}^{2}+\epsilon_{n}^{2}\right)+o(1). \end{flalign*}	
	The stated result then follows from Markov's inequality,
	\begin{equation*}
	\E_{P_{0}}^{} \widehat{Q}\left(d(\theta,\theta_\star)>M_{n} n\left(\epsilon_{n}^{2}+\kappa_{n}^{2}\right)\right) \leq \frac{\E_{P_0}^{} \widehat{Q}\left\{d(\theta,\theta_\star)\right\}}{M_{n} n\left(\epsilon_{n}^{2}+\kappa_{n}^{2}\right)} \lesssim M_n^{-1} \rightarrow 0.
	\end{equation*}

\end{proof}

\subsection{Proofs of Theorems in Section \ref{theory}}

Using Lemmas \ref{lemma:conv} and \ref{thm:one}, we can now prove Theorems %
\ref{thm:merge1} and \ref{thm:merge} stated in the main text. A simple proof
of Theorem \ref{thm:merge1} can be given by first proving Theorem \ref%
{thm:merge}. Therefore, we first prove Theorem \ref{thm:merge}. 

	\bigskip

\begin{proof}{\bf of Theorem \ref{thm:merge}.}
	For probability measures $P$ and $Q$, let $d_{H}(P,Q)$ denote the Hellinger distance between $P$ and $Q$. Consider a positive sequence $\epsilon_n>0$, and define the set $A(\epsilon_n):=\{P^{(n)}_\theta\in\mathcal{P},\;\theta\in\Theta:d_{H}^2(P^{(n)}_\theta,P_\star^{(n)})\ge\epsilon_n\}$. By convexity of $d_H^2(\cdot,\cdot)$ in the first argument and Jensen's inequality, 
	\begin{flalign*}
	d^2_H(P^{(n)}_Q,P^{(n)}_\star)&\leq \int_{\Theta} d_H^2(P^{(n)}_\theta,P^{(n)}_\star)\dt\widehat{Q}(\theta)
	\\
	&= \int_{A^c(\epsilon^2_n)} d_H^2(P^{(n)}_\theta,P^{(n)}_\star)\dt\widehat{Q}(\theta)+\int_{A(\epsilon^2_n)} d_H^2(P^{(n)}_\theta,P^{(n)}_\star)\dt\widehat{Q}(\theta)\\&\leq\epsilon^2_n +\sqrt{2}\widehat{Q}\{ A(\epsilon^2_n)\}.
	\end{flalign*}By Lemma \ref{thm:one}, for any $\epsilon_n$ such that $n\epsilon_n^2\rightarrow\infty$, $\widehat{Q}\{A(\epsilon_n)\}=o_{p}(1)$, so that $d^2_H(P^{(n)}_Q,P^{(n)}_\star)\le \epsilon^2_n$  (wpc1). Using the above and the definition $
	d_H(P^{(n)}_Q,P^{(n)}_\star)=\sqrt{d^2_H(P^{(n)}_Q,P^{(n)}_\star)},
	$
	we can conclude that, with probability converging to one, $$
	d_H(P^{(n)}_Q,P^{(n)}_\star)\le\epsilon_n.
	$$Which yields the first stated result. Applying the relationship between total variation and Hellinger distance yields the result: $
	d_{TV}(P^{(n)}_Q,P^{(n)}_{\Pi_{w_n}})\leq \sqrt{2}d_H(P^{(n)}_Q,P^{(n)}_{\Pi_{w_n}}).
	$
\end{proof}


\begin{proof}{\bf of  Theorem \ref{thm:merge1}.}
	A similar argument to the proof of Theorem \ref{thm:merge1} yields 
	\begin{flalign*}
	d^2_H(P^{(n)}_{\Pi_{w_n}},P^{(n)}_\star)&\leq \int_{\Theta} d_H^2(P^{(n)}_{\theta},P^{(n)}_\star)\dt\Pi_{w_n}(\theta|\y)\leq\epsilon^2_n+\sqrt{2}\Pi_{w_n}\{A(\epsilon^2_n)|\y\}\leq \epsilon^2_n,
	\end{flalign*}
	where the last line follows from the convergence in Lemma \ref{lemma:conv} and holds wpc1. Apply the triangle inequality, and the relationship between the Hellinger and it's square, to see that (wpc1)
	\begin{flalign*}
	d_H(P^{(n)}_Q,P^{(n)}_{\Pi_{w_n}})&\leq d_H(P^{(n)}_Q,P^{(n)}_\star)+d_H(P^{(n)}_{\theta},P^{(n)}_\star)\\&= \sqrt{d^2_H(P^{(n)}_\theta,P^{(n)}_\star)}+\sqrt{d^2_H(P^{(n)}_\theta,P^{(n)}_\star)}\\&\le 2\epsilon_n.
	\end{flalign*}The above holds for any $\epsilon_n\rightarrow0$ with $n\epsilon_n^2\rightarrow0$, and we can conclude that $d_{TV}(P^{(n)}_{Q},P^{(n)}_{\Pi_{w_n}})\le \sqrt{2}d_H(P^{(n)}_{Q},P^{(n)}_{\Pi_{w_n}})=o_p(1)$. 
	
\end{proof}

\subsection{Additional Lemmas}\label{app:replems}
The following Lemmas from \cite{zhang2020convergence} are used in the proof of Lemma \ref{thm:one}, and are reproduce here, without proof, to aid the reader. We again use $\E_{P_0}$ to denote expectations taken under $P_0$. 

\begin{lem}\label{lem:general1}
	For $\widehat{Q}$  in Definition \ref{def:two}, we have
	\begin{eqnarray*}
		&& \E_{P_0}^{}\widehat QL(P_\theta^{(n)}, P_0^{(n)}) \\
		&\leq& \inf_{a>0}\frac{1}{a}\left(\inf_{Q\in\mathcal{S}}\E_{P_0}^{}D( Q\|\Pi(\cdot|X^{(n)}))+\log \E_{P_0}^{}\Pi(\exp(aL(P_\theta^{(n)}, P_0^{(n)}))|X^{(n)})\right).
	\end{eqnarray*}
\end{lem}

\begin{lem}\label{lem:sub-exp}
	Suppose the random variable $X$ satisfies
	\[\mathbb{P}(X\geq t)\leq c_1\exp(-c_2t),\]
	for all $t\geq t_0>0$.
	Then, for any $0<a\leq \frac{1}{2}c_2$,
	\[\mathbb{E}\exp(aX)\leq \exp(at_0)+c_1.\]
\end{lem}

\bibliography{Prediction_bib_LBVB}

\newpage 

\begin{center}
	{\Large\bf Supplementary material for ``Loss-Based Variational Bayes Prediction''}
\end{center}

\setcounter{page}{1}
\setcounter{section}{0}
\setcounter{table}{0}
\setcounter{figure}{0}
\setcounter{equation}{0}
\setcounter{rem}{1} 
\renewcommand{\theHsection}{SIsection.\arabic{section}}
\renewcommand{\theHtable}{SItable.\arabic{table}}
\renewcommand{\theHfigure}{SIfigure.\arabic{figure}}
\renewcommand{\theHequation}{SIequation.\arabic{section}.\arabic{equation}}
\renewcommand{\thepage}{S\arabic{page}}  
\renewcommand{\thesection}{S\arabic{section}}   
\renewcommand{\thetable}{S\arabic{table}}   
\renewcommand{\thefigure}{S\arabic{figure}}
\renewcommand{\theequation}{S\arabic{section}.\arabic{equation}}

\section{Details for the Implementation of all Variational Approximations}

\subsection{General Notational Matters}

\label{App:Gradients1} Throughout this appendix we will employ the following
notation. All gradients are column vectors and their notation starts with
the symbol $\nabla $. For two generic matrices $A_{d_{1}\times d_{2}}$ and $%
B_{d_{3}\times d_{4}}$ we have that 
\begin{equation*}
\frac{\partial A}{\partial B}=\frac{\partial \text{vec}(A)}{\partial \text{%
		vec}(B)},
\end{equation*}%
where vec is the vectorization operation and $\frac{\partial A}{\partial B}$
is a matrix of dimension $(d_{1}d_{2})\times (d_{3}d_{4})$. Throughout this
appendix scalars are treated as matrices of dimension $1\times 1$.

The gradient of the log density of the approximation is computed as 
\begin{equation*}
\nabla _{\theta }\log q_{\lambda }\left( {\theta }\right) =-\left(
D^{2}\right) ^{-1}\left( {\theta }-{\mu }\right) .
\end{equation*}%
To compute $\frac{\partial {\theta }}{\partial {\lambda }}$ note that 
\begin{equation*}
\frac{\partial {\theta }}{\partial {\lambda }}=\left[ \frac{\partial {\theta 
}}{\partial {\mu }},\frac{\partial {\theta }}{\partial {d}}\right] ,
\end{equation*}%
where 
\begin{equation*}
\frac{\partial {\theta }}{\partial {\mu }}=I_{r},\ \ \ \ \ \ \ \frac{%
	\partial {\theta }}{\partial {d}}=\text{diag}({e}).
\end{equation*}

\section{Stochastic Gradient Ascent}

\label{sga} The optimization problem in \eqref{Eq:optimization} is performed
via stochastic gradient ascent methods (SGA). SGA maximizes $\mathcal{L}%
(\lambda )$ by first initializing the variational parameters at some vector
of values ${\lambda }^{(0)}$, and then sequentially iterating over 
\begin{equation*}
{\lambda }^{i+1}={\lambda }^{i}+\Delta {\lambda }^{i+1}\left[ \widehat{
	\nabla _{\lambda }\mathcal{L}\left( {\lambda }^{(i)}\right) },{\rho }\right]
.
\end{equation*}%
The step size $\Delta {\lambda }^{i+1}$ is a function of an unbiased
estimate of the ELBO gradient, $\widehat{\nabla _{\lambda }\mathcal{L}\left( 
	{\lambda }^{(i)}\right) }$, and the set\textbf{\ }of tuning parameters that
control the learning rates in the optimization problem, denoted by ${\rho }$%
. Throughout this paper we employ the ADADELTA method of Zieler (2012) to
update $\Delta {\lambda }^{i+1}$.

Key to achieving\ \textit{fast} maximization of $\mathcal{L}(\lambda )$ via
SGA is the use of a variance-reduction technique in producing the unbiased
gradient estimate $\widehat{\nabla _{\lambda }\mathcal{L}\left( {\lambda }
	^{(i)}\right) }$. Here, we follow Kingma and Welling (2014) and Rezende et
al. (2014), and make use of the `reparameterization trick'. In this
approach, a draw ${\theta }$ from $q_{\lambda }$ is written as a direct
function of ${\lambda }$, and a set of random variables ${\varepsilon }$
that are invariant with respect to ${\lambda }$. For the {mean-field}
approximation used for this example we can write ${\theta }\left( {\lambda },%
{\varepsilon }\right) =\mu +{d}\circ {\varepsilon }$, with ${\varepsilon }%
\sim N\left( {0}_{4},I_{4}\right) $. This reparametrization allows us to
re-write $\mathcal{L}(\lambda )$ as 
\begin{equation}
\mathcal{L}(\lambda )=\mathbb{E}_{{\varepsilon }}\left[ wS_{n}^{j}\left( {\
	\theta }\left( {\lambda },{\varepsilon }\right) \right) +\log \pi \left( {\
	\theta }\left( {\lambda },{\varepsilon }\right) \right) -\log q_{\lambda
}\left( {\theta }\left( {\lambda },{\varepsilon }\right) \right) \right]
\label{Eq:ELBOrepar}
\end{equation}%
and its gradient as 
\begin{equation}
\nabla _{\lambda }\mathcal{L}(\lambda )=\mathbb{E}_{{\varepsilon }}\left( 
\frac{\partial {\theta }\left( {\lambda },{\varepsilon }\right) }{\partial {%
		\ \lambda }}^{\prime }\left[ w\nabla _{\theta }S_{n}^{j}\left( {\theta }%
\left( {\lambda },{\varepsilon }\right) \right) +\nabla _{\theta }\log \pi
\left( {\ \theta }\left( {\lambda },{\varepsilon }\right) \right) -\nabla
_{\theta }\log q_{\lambda }\left( {\theta }\left( {\lambda },{\varepsilon }%
\right) \right) \right] \right) .  \label{Eq:ELBOrepargrad}
\end{equation}%
A low-variance unbiased estimate of $\nabla _{\lambda }\mathcal{L}(\lambda )$
can be constructed by numerically computing the expectation in (\ref%
{Eq:ELBOrepargrad}) using the (available) closed-form expressions for the
derivatives $\frac{\partial {\theta }\left( {\lambda },{\varepsilon }\right) 
}{\partial {\lambda }}$, $\nabla _{\theta }\log q_{\lambda }\left( {\theta }%
\left( {\lambda },{\varepsilon }\right) \right) $, $\nabla _{\theta }\log
\pi \left( {\theta }\left( {\lambda },{\varepsilon }\right) \right) $, and $%
\nabla _{\theta }S_{n}^{j}\left( {\theta }\left( {\lambda },{\varepsilon }%
\right) \right) $ for any $j$. We follow \cite{kingma2019introduction} and
use a single draw of $\varepsilon $ for the construction of an estimate of %
\eqref{Eq:ELBOrepargrad}. The required expressions for the GARCH model are
provided in Appendix \ref{garch}.

\section{The GARCH Predictive Class\label{garch} (Section \protect\ref%
	{sec:toy})}

The predictive class, $\mathcal{P}^{(t)}$, is defined by a generalized
autoregressive conditional heteroscedastic GARCH(1,1) model with Gaussian
errors, $Y_{t}=\theta _{1}^{r}+\sigma _{t}\varepsilon _{t},$\ $\varepsilon
_{t}\overset{i.i.d.}{\sim}N\left( 0,1\right) ,$\ $\sigma _{t}^{2}=\theta
_{2}^{r}+\theta _{3}^{r}\left( Y_{t-1}-\theta _{1}^{r}\right) ^{2}+\theta
_{4}^{r}\sigma _{t-1}^{2},$\ with $\theta =\left( \theta _{1},\theta
_{2},\theta _{3},\theta _{4}\right) ^{\prime }=\left( \theta _{1}^{r},\log
(\theta _{2}^{r}),\Phi _{1}^{-1}\left( \theta _{3}^{r}\right) ,\Phi
_{1}^{-1}\left( \theta _{4}^{r}\right) \right) ^{\prime }$. Note that
throughout this section $\theta _{1}$ and $\theta _{1}^{r}$ can be used
interchangeably.

\subsection{Priors}

We employ the following priors for each of the parameters of the model: 
\begin{equation*}
p(\theta _{1})\propto 1,\hspace{1cm}p({\theta _{2}^{r}})\propto \frac{1}{{%
		\theta _{2}^{r}}}I({\theta _{2}^{r}}>0),\hspace{1cm}{\theta _{3}^{r}}\sim
U(0,1)\text{,\ \ \ and }\hspace{0.5cm}{\theta _{4}^{r}}\sim U(0,1).
\end{equation*}%
For the implementation of variational inference, all the parameters are
transformed into the real line as follows: 
\begin{align*}
& \text{(i) }{\theta _{2}^{r}}\text{ is transformed to }{\theta _{2}}=\log ({%
	\theta _{2}^{r}});\hspace{1cm}\text{(ii) }{\theta _{3}^{r}}\text{ is
	transformed to }{\theta _{3}}=\Phi _{1}^{-1}\left( {\theta _{3}^{r}}\right) ;
\\
& \text{(iii) }{\theta _{4}^{r}}\text{ is transformed to }{\theta _{4}}=\Phi
_{1}^{-1}\left( {\theta _{4}^{r}}\right) .
\end{align*}%
After applying these transformations, we have that the prior densities are: 
\begin{equation*}
\text{(i) }p({\theta _{1}})\propto 1;\hspace{1cm}\text{(ii) }p({\theta _{2}}%
)\propto 1;\hspace{1cm}\text{(iii) }p({\theta _{3}})=\phi _{1}\left( {\theta
	_{3}}\right) ;\hspace{0.65cm}\text{(iv) }p({\theta _{4}})=\phi _{1}\left( {%
	\theta _{4}}\right) .
\end{equation*}%
The gradient of the logarithm of the prior is $\nabla _{\theta }\log p\left( 
{\theta }\right) =\left( 0,0,-{\theta _{3}},-\theta _{4}\right) ^{^{\prime
}} $.

\subsection{Derivation of \texorpdfstring{$\nabla_\theta
		S_{n}\left({%
			\theta}\right)$}{TEXT} for all scoring rules}

We can show that $\nabla _{\theta }S_{n}\left( {\theta }\right)
=\sum_{t=1}^{n}\nabla _{\theta }s\left( P_{{\theta }}^{(t-1)},y_{t}\right) .$
Thus, we must find an expression for $\nabla _{\theta }s\left( P_{{\theta }%
}^{(t-1)},y_{t}\right) $ for each of the scores. The gradients from all the
scores can be written as a function of the recursive derivatives: 
\begin{align*}
\text{(i)}\ \ \frac{\partial \sigma _{t}^{2}}{\partial {\theta _{1}}}=&
-2\theta _{3}^{r}(y_{t-1}-{\theta _{1}})+\theta _{4}^{r}\frac{\partial
	\sigma _{t-1}^{2}}{\partial {\theta _{1}}};\hspace{1cm}\text{(ii)}\ \ \frac{%
	\partial \sigma _{t}^{2}}{\partial \theta _{2}^{r}}=1+\theta _{4}^{r}\frac{%
	\partial \sigma _{t-1}^{2}}{\partial \theta _{2}^{r}}; \\
\text{(iii)}\ \ \frac{\partial \sigma _{t}^{2}}{\partial \theta _{3}^{r}}=&
(y_{t-1}-{\theta _{1}})^{2}+\theta _{4}^{r}\frac{\partial \sigma _{t-1}^{2}}{%
	\partial \theta _{3}^{r}};\hspace{1.9cm}\text{(iv)}\ \ \frac{\partial \sigma
	_{t}^{2}}{\partial \theta _{4}^{r}}=\theta _{4}^{r}\frac{\partial \sigma
	_{t-1}^{2}}{\partial \theta _{4}^{r}}+\sigma _{t-1}^{2}.
\end{align*}%
with $\frac{\partial \sigma _{0}^{2}}{\partial {\theta _{1}}}=0$, $\frac{%
	\partial \sigma _{0}^{2}}{\partial \theta _{2}^{r}}=0$, $\frac{\partial
	\sigma _{0}^{2}}{\partial \theta _{3}^{r}}=0$ and $\frac{\partial \sigma
	_{0}^{2}}{\partial \theta _{4}^{r}}=0$.

\subsubsection{Logarithmic score (LS)}

The gradient for the LS can be written as 
\begin{equation*}
\nabla _{\theta }s^{\text{LS}}\left( P_{{\theta }}^{(t-1)},y_{t}\right)
=\left( \nabla _{{\theta _{1}}}s^{\text{LS}}\left( P_{{\theta }%
}^{(t-1)},y_{t}\right) ^{\prime },\nabla _{\theta _{2}}s^{\text{LS}}\left(
P_{{\theta }}^{t-1},y_{t}\right) ,\nabla _{\theta _{3}}s^{\text{LS}}\left(
P_{{\theta }}^{t-1},y_{t}\right) ,\nabla _{\theta _{4}}s^{\text{LS}}\left(
P_{{\theta }}^{t-1},y_{t}\right) \right) ^{\prime },
\end{equation*}%
with {\footnotesize 
	\begin{align*}
	\nabla _{{\theta _{1}}}S^{\text{LS}}\left( P_{{\theta }}^{(t-1)},y_{t}%
	\right) =& -\frac{1}{2\sigma _{t}^{2}}\frac{\partial \sigma _{t}^{2}}{%
		\partial {\theta _{1}}}+\frac{1}{2}\frac{(y_{t}-{\theta _{1}})^{2}}{\sigma
		_{t}^{4}}\frac{\partial \sigma _{t}^{2}}{\partial {\theta _{1}}}+\frac{%
		(y_{t}-{\theta _{1}})}{\sigma _{t}^{2}},\hspace{0.3cm}\nabla _{\theta
		_{2}}S^{\text{LS}}\left( P_{{\theta }}^{t-1},y_{t}\right) =\theta _{2}^{r}%
	\left[ -\frac{1}{2\sigma _{t}^{2}}\frac{\partial \sigma _{t}^{2}}{\partial
		\theta _{2}^{r}}+\frac{1}{2}\frac{(y_{t}-{\theta _{1}})^{2}}{\sigma _{t}^{4}}%
	\frac{\partial \sigma _{t}^{2}}{\partial \theta _{2}^{r}}\right] , \\
	\nabla _{\theta _{3}}S^{\text{LS}}\left( P_{{\theta }}^{(t-1)},y_{t}\right)
	=& \phi _{1}(\theta _{3})\left[ -\frac{1}{2\sigma _{t}^{2}}\frac{\partial
		\sigma _{t}^{2}}{\partial \theta _{3}^{r}}+\frac{1}{2}\frac{(y_{t}-{\theta
			_{1}})^{2}}{\sigma _{t}^{4}}\frac{\partial \sigma _{t}^{2}}{\partial \theta
		_{3}^{r}}\right] \hspace{0.3cm}\text{and}\hspace{0.3cm}\nabla _{\theta
		_{4}}S^{\text{LS}}\left( P_{{\theta }}^{(t-1)},y_{t}\right) =\phi
	_{1}(\theta _{4})\left[ -\frac{1}{2\sigma _{t}^{2}}\frac{\partial \sigma
		_{t}^{2}}{\partial \theta _{4}^{r}}+\frac{1}{2}\frac{(y_{t}-{\theta _{1}}%
		)^{2}}{\sigma _{t}^{4}}\frac{\partial \sigma _{t}^{2}}{\partial \theta
		_{4}^{r}}\right] .
	\end{align*}%
}

\subsubsection{Continuously ranked probability score (CRPS)}

For the Gaussian GARCH(1,1)\textbf{\ }predictive class the CRPS can be
expressed as 
\begin{equation*}
s^{\text{CRPS}}\left( P_{{\theta }}^{(t-1)},y_{t}\right) =-\sigma _{t}B_{t},
\end{equation*}%
with $B_{t}=z_{t}\left( 2\Phi _{1}(z_{t})-1\right) +2\phi _{1}(z_{t})-\frac{1%
}{\sqrt{\pi }}$ and $z_{t}=\frac{y_{t}-{\theta _{1}}}{\sigma _{t}}$. The
gradient of the CRPS can be written as 
\begin{equation*}
\nabla _{\theta }s^{\text{CRPS}}\left( P_{{\theta }}^{(t-1)},y_{t}\right)
=\left( \nabla _{{\theta _{1}}}s^{\text{CRPS}}\left( P_{{\theta }%
}^{t-1},y_{t}\right) ,\nabla _{{\theta _{2}}}s^{\text{CRPS}}\left( P_{{%
		\theta }}^{t-1},y_{t}\right) ,\nabla _{{\theta _{3}}}s^{\text{CRPS}}\left(
P_{{\theta }}^{t-1},y_{t}\right) ,\nabla _{{\theta _{4}}}s^{\text{CRPS}%
}\left( P_{{\theta }}^{t-1},y_{t}\right) \right) ^{\prime }.
\end{equation*}%
The elements of this gradient are given by {\footnotesize \ 
	\begin{align}
	\nabla _{{\theta _{1}}}s^{\text{CRPS}}\left( P_{{\theta }}^{(t-1)},y_{t}%
	\right) & =-\sigma _{t}\frac{\partial B_{t}}{\partial z_{t}}\left[ \frac{%
		\partial z_{t}}{\partial \sigma _{t}^{2}}\frac{\partial \sigma _{t}^{2}}{%
		\partial {\theta _{1}}}-\frac{1}{\sigma _{t}}\right] -\frac{B_{t}}{2\sigma
		_{t}}\frac{\partial \sigma _{t}^{2}}{\partial {\theta _{1}}}  \notag \\
	\nabla _{{\theta _{2}}}s^{\text{CRPS}}\left( P_{{\theta }}^{(t-1)},y_{t}%
	\right) & =-\sigma _{t}\frac{\partial B_{t}}{\partial z_{t}}\frac{\partial
		z_{t}}{\partial \sigma _{t}^{2}}\frac{\partial \sigma _{t}^{2}}{\partial
		\theta _{2}^{r}}-\frac{B_{t}}{2\sigma _{t}}\frac{\partial \sigma _{t}^{2}}{{%
			\theta _{2}^{r}}}{\theta _{2}^{r}}  \notag \\
	\nabla _{{\theta _{3}}}s^{\text{CRPS}}\left( P_{{\theta }}^{(t-1)},y_{t}%
	\right) & =-\sigma _{t}\frac{\partial B_{t}}{\partial z_{t}}\frac{\partial
		z_{t}}{\partial \sigma _{t}^{2}}\frac{\partial \sigma _{t}^{2}}{\partial {%
			\theta _{3}^{r}}}-\frac{B_{t}}{2\sigma _{t}}\frac{\partial \sigma _{t}^{2}}{{%
			\theta _{3}^{r}}}\phi _{1}({\theta _{3}})  \notag \\
	\nabla _{{\theta _{4}}}s^{\text{CRPS}}\left( P_{{\theta }}^{(t-1)},y_{t}%
	\right) & =-\sigma _{t}\frac{\partial B_{t}}{\partial z_{t}}\frac{\partial
		z_{t}}{\partial \sigma _{t}^{2}}\frac{\partial \sigma _{t}^{2}}{\partial
		\theta _{4}^{r}}-\frac{B_{t}}{2\sigma _{t}}\frac{\partial \sigma _{t}^{2}}{{%
			\theta _{4}^{r}}}\phi _{1}({\theta _{4}})  \notag
	\end{align}%
} with $\frac{\partial B_{t}}{\partial z_{t}}=2\Phi _{1}(z_{t})-1$ and $%
\frac{\partial z_{t}}{\partial \sigma _{t}^{2}}=-\frac{z_{t}}{2\sigma
	_{t}^{2}}$.

\subsubsection{Censored logarithmic score (CLS)}

For some threshold value $y_{q}$, denote the upper tail support of
predictive distribution as $A=\{y_{t}:y_{t}>y_{q}\}$. The gradient for the
upper tail CLS can be written as 
\begin{eqnarray*}
	&&\nabla _{\theta }s^{\text{CLS-A}}\left( P_{{\theta }}^{(t-1)},y_{t}\right)
	\\
	&=&\left( \nabla _{{\theta _{1}}}s^{\text{CLS-A}}\left( P_{{\theta }%
	}^{(t-1)},y_{t}\right) ,\nabla _{{\theta _{2}}}s^{\text{CLS-A}}\left( P_{{%
			\theta }}^{(t-1)},y_{t}\right) ,\nabla _{{\theta _{3}}}s^{\text{CLS-A}%
	}\left( P_{{\theta }}^{(t-1)},y_{t}\right) ,\nabla _{\theta _{4}}s^{\text{%
			CLS-A}}\left( P_{{\theta }}^{(t-1)},y_{t}\right) \right) ^{\prime },
\end{eqnarray*}%
with {\footnotesize 
	\begin{align}
	\nabla _{{\theta _{1}}}s^{\text{CLS-A}}\left( P_{{\theta }%
	}^{(t-1)},y_{t}\right) & =\nabla _{{\theta _{1}}}s^{\text{LS}}\left( P_{{%
			\theta }}^{(t-1)},y_{t}\right) I(y_{t}\in A)+\frac{\phi _{1}(\frac{y_{q}-{%
				\theta _{1}}}{\sigma _{t}})}{P_{{\theta }}^{(t-1)}(y_{q})}\left( -\frac{%
		y_{q}-{\theta _{1}}}{2\sigma _{t}^{3}}\frac{\partial \sigma _{t}^{2}}{%
		\partial {\theta _{1}}}-\frac{1}{\sigma _{t}}\right) I(y_{t}\in A^{c}) 
	\notag \\
	\nabla _{{\theta _{2}}}s^{\text{CLS-A}}\left( P_{{\theta }%
	}^{(t-1)},y_{t}\right) & =\nabla _{{\theta _{2}}}s^{\text{LS}}\left( P_{{%
			\theta }}^{(t-1)},y_{t}\right) I(y_{t}\in A)+\frac{\phi _{1}(\frac{%
			y_{q}-\theta _{1}}{\sigma _{t}})}{P_{{\theta }}^{(t-1)}(y_{q})}\left( -\frac{%
		y_{q}-{\theta _{1}}}{2\sigma _{t}^{3}}\frac{\partial \sigma _{t}^{2}}{%
		\partial \theta _{2}^{r}}\right) {\theta _{2}^{r}}I(y_{t}\in A^{c})  \notag
	\\
	\nabla _{{\theta _{3}}}s^{\text{CLS-A}}\left( P_{{\theta }%
	}^{(t-1)},y_{t}\right) & =\nabla _{{\theta _{3}}}s^{\text{LS}}\left( P_{{%
			\theta }}^{(t-1)},y_{t}\right) I(y_{t}\in A)+\frac{\phi _{1}(\frac{%
			y_{q}-\theta _{1}}{\sigma _{t}})}{P_{{\theta }}^{(t-1)}(y_{q})}\left( -\frac{%
		y_{q}-{\theta _{1}}}{2\sigma _{t}^{3}}\frac{\partial \sigma _{t}^{2}}{%
		\partial \theta _{3}^{r}}\right) \phi _{1}({\theta _{3}})I(y_{t}\in A^{c}) 
	\notag \\
	\nabla _{{\theta _{4}}}s^{\text{CLS-A}}\left( P_{{\theta }%
	}^{(t-1)},y_{t}\right) & =\nabla _{{\theta _{4}}}s^{\text{LS}}\left( P_{{%
			\theta }}^{(t-1)},y_{t}\right) I(y_{t}\in A)+\frac{\phi _{1}(\frac{%
			y_{q}-\theta _{1}}{\sigma _{t}})}{P_{{\theta }}^{(t-1)}(y_{q})}\left( -\frac{%
		y_{q}-{\theta _{1}}}{2\sigma _{t}^{3}}\frac{\partial \sigma _{t}^{2}}{%
		\partial \theta _{4}^{r}}\right) \phi _{1}({\theta _{4}})I(y_{t}\in A^{c}) 
	\notag
	\end{align}%
} To obtain the gradient expressions for the lower tail CLS simply replace
the term $\frac{\phi _{1}(\frac{y_{q}-{\theta _{1}}}{\sigma _{t}})}{P_{{%
			\theta }}^{(t-1)}(y_{q})}$ by the term $-\frac{\phi _{1}(\frac{y_{q}-{\theta
			_{1}}}{\sigma _{t}})}{1-P_{{\theta }}^{(t-1)}(y_{q})}$, and redefine the set 
$A$.

\subsubsection{Interval score (IS)}

The gradient of the IS can be written as {\small 
	\begin{equation*}
	\nabla _{\theta }s^{\text{IS}}\left( P_{{\theta }}^{(t-1)},y_{t}\right)
	=\left( \nabla _{{\theta _{1}}}s^{\text{IS}}\left( P_{{\theta }%
	}^{(t-1)},y_{t}\right) ,\nabla _{{\theta _{2}}}s^{\text{IS}}\left( P_{{%
			\theta }}^{(t-1)},y_{t}\right) ,\nabla _{{\theta _{3}}}s^{\text{IS}}\left(
	P_{{\theta }}^{(t-1)},y_{t}\right) ,\nabla _{{\theta _{4}}}s^{\text{IS}%
	}\left( P_{{\theta }}^{(t-1)},y_{t}\right) \right) .
	\end{equation*}%
} For $\theta _{i}\in \{{\theta _{1}},{\theta _{2}},{\theta _{3}},\theta
_{4}\}$, the elements of the gradient can be written as 
\begin{equation*}
\nabla _{\theta _{i}}s^{\text{IS}}\left( P_{{\theta }}^{(t-1)},y_{t}%
\right) =-\left[ 1-\frac{2}{1-q}I(y_{t}>u_{t})\right] \frac{\partial u_{t}}{%
	\partial \theta _{i}}-\left[ \frac{2}{1-q}I(y_{t}<l_{t})-1\right] \frac{%
	\partial l_{t}}{\partial \theta _{i}},
\end{equation*}%
where the derivative $\frac{\partial u_{t}}{\partial {\theta _{i}}}$ can be
evaluated as $\frac{\partial u_{t}}{\partial \theta _{i}}=-\frac{\partial
	u_{t}}{\partial \alpha _{u,t}}\frac{\partial \alpha _{u,t}}{\partial \theta
	_{i}}$ with $\alpha _{u,t}=P_{{\theta }}^{(t-1)}(u_{t})$. The first term can
be computed as $\frac{\partial u_{t}}{\partial \alpha _{u,t}}=\frac{1}{%
	p\left( u_{t}|y_{1:t-1},{\theta }\right) }$. The second term is parameter
specific, and can be computed as

\begin{align*}
\text{(i) }\ \ \frac{\partial \alpha_{u,t}}{\partial {\theta_1}} &
=\phi_1\left(\frac{u_t-{\theta_1}}{\sigma_t}\right)\left[\left(-\frac{%
	u_t-\theta_1}{2\sigma_t^3}\right)\frac{\partial\sigma_t^2}{\partial{\theta_1}%
} -\frac{1}{\sigma_t}\right];\hspace{0.5cm}\text{(ii) }\ \ \frac{\partial
	\alpha_{u,t}}{\partial {\theta_2}} = \phi_1\left(\frac{u_t-{\theta_1}}{%
	\sigma_t}\right)\left(-\frac{u_t-{\theta_1}}{2\sigma_t^3}\right)\frac{%
	\partial\sigma_t^2}{\partial{\theta_2^r}}{\theta_2^r}; \\
\text{(iii) }\ \ \frac{\partial \alpha_{u,t}}{\partial {\theta_3}} &
=\phi_1\left(\frac{u_t-{\theta_1}}{\sigma_t}\right)\left(-\frac{u_t-{\theta_1%
}}{2\sigma_t^3}\right)\frac{\partial\sigma_t^2}{\partial\theta_3^r}\phi_1({%
	\theta_3});\hspace{0.6cm}\text{(iv) }\ \ \frac{\partial\alpha_{u,t}}{%
	\partial {\theta_4}} =\phi_1\left(\frac{u_t-{\theta_1}}{\sigma_t}%
\right)\left(-\frac{u_t-{\theta_1}}{2\sigma_t^3}\right)\frac{%
	\partial\sigma_t^2}{\partial{\theta_4^r}}\phi_1({\theta_4}).
\end{align*}

The derivative $\frac{\partial l_{t}}{\partial {\theta _{i}}}$ is evaluated
in the same fashion as described for $\frac{\partial u_{t}}{\partial {\theta
		_{i}}}$.

\section{Autoregressive Mixture Predictive Class (Section \protect\ref{ar})}

\label{App:priorchoice}

\subsection{Priors}

We employ the following priors for each of the parameters of the model: 
\begin{equation*}
\beta _{k,0}\sim N(0,10000^{2}),\hspace{0.5cm}\beta _{k,1}\sim U(-1,1),%
\hspace{0.5cm}v_{k}\sim \text{Beta}(1,2),\hspace{0.5cm}\left( \sigma
_{k}^{2}\right) ^{-1}\sim \mathcal{G}(1,1)\text{,\ \ \ and }\hspace{0.5cm}%
\mu \sim N(0,100^{2}),
\end{equation*}%
where the parameter $v_{k}$ is given by the stick breaking decomposition of
the mixture weights, which sets $\tau _{k}=v_{k}\prod_{j=1}^{k-1}(1-v_{j})$,
for $k=1,\dots ,K$ and where $v_{K}=1$. For the implementation of
variational inference, all the parameters are transformed into the real line
as follows: 
\begin{align*}
& \text{(i) }\beta _{k,1}\text{ is transformed to }\eta _{k}=\Phi
_{1}^{-1}\left( \frac{1}{2}\left( \beta _{k,1}+1\right) \right) ;\hspace{1cm}%
\text{(ii) }v_{k}\text{ is transformed to }\psi _{k}=\Phi _{1}^{-1}\left(
v_{k}\right) ; \\
& \text{(iii) }\sigma _{k}\text{ is transformed to }\kappa _{k}=\log \sigma
_{k}.
\end{align*}%
After applying these transformations, we have that the prior densities are: 
\begin{align*}
& \text{(i) }p(\beta _{k,0})=\phi _{1}\left( \beta _{k,1};0,10000^{2}\right)
;\hspace{1cm}\text{(ii) }p(\eta _{k})=\phi _{1}\left( \eta _{k};0,1\right) ;
\\
& \text{(iii) }p(\psi _{k})\propto \left[ 1-\Phi _{1}\left( \psi _{k}\right) %
\right] \phi _{1}\left( \psi _{k}\right) ;\hspace{0.65cm}\text{(iv) }%
p(\kappa _{k})\propto 2\exp \left( -2\kappa _{k}\right) \exp \left[ -\exp
\left( -2\kappa _{k}\right) \right] ; \\
& \text{(v) }p(\mu )=\phi _{1}\left( \mu ;0,100^{2}\right) .
\end{align*}%
\textbf{Computation of 
	\texorpdfstring{$\nabla_\theta \log
		p\left({\theta}\right)$}{TEXT}}\label{App:Gradients2}\medskip \newline
Denoting as ${\beta }_{0}=\left( \beta _{1,0},\dots ,\beta _{K,0}\right)
^{\prime }$, ${\eta }=\left( \eta _{1},\dots ,\eta _{K}\right) ^{\prime }$, $%
{\psi }=\left( \psi _{1},\dots ,\psi _{K}\right) ^{\prime }$ and ${\kappa }%
=\left( \kappa _{1},\dots ,\kappa _{K}\right) ^{\prime }$, the gradient of
the prior density is 
\begin{equation*}
\nabla _{\theta }\log p\left( {\theta }\right) =\left[ \frac{\partial \log
	p\left( {\theta }\right) }{\partial {\beta }_{0}},\frac{\partial \log
	p\left( {\theta }\right) }{\partial {\eta }},\frac{\partial \log p\left( {%
		\theta }\right) }{\partial {\psi }},\frac{\partial \log p\left( {\theta }%
	\right) }{\partial {\kappa }},\frac{\partial \log p\left( {\theta }\right) }{%
	\mu }\right] ^{\prime },
\end{equation*}%
with%
\begin{equation*}
\text{(i) }\frac{\partial \log p\left( {\theta }\right) }{\partial {\beta }%
	_{0}}=-(10000)^{-2}{\beta }_{0}^{^{\prime }};\ \ \ \ \ \text{(ii) }\frac{%
	\partial \log p\left( {\theta }\right) }{\partial {\eta }}=-{\eta }%
^{^{\prime }};
\end{equation*}%
\begin{equation*}
\text{(iii) }\frac{\partial \log p\left( {\theta }\right) }{\partial {\psi }}%
=-(1-{v}^{\prime })^{-1}\frac{\partial {v}}{\partial \psi }-{\psi }%
^{^{\prime }};\ \ \ \ \ \text{(iv) }\frac{\partial \log p\left( {\theta }%
	\right) }{\partial {\kappa }}=2\tilde{{\kappa }}^{^{\prime }}-2;
\end{equation*}%
\begin{equation*}
\text{(v) }\frac{\partial \log p\left( {\theta }\right) }{\partial {\mu }}%
=-(100)^{-2}{\mu }.
\end{equation*}%
and where $(1-{v}^{\prime })^{-1}=\left( [1-v_{1}]^{-1},\dots
,[1-v_{K-1}]^{-1}\right) $ and $\tilde{{\kappa }}=(\exp [-2\kappa
_{1}],\dots ,\exp [-2\kappa _{K}])^{\prime }$.

\subsection{Derivation of \texorpdfstring{$\nabla_\theta
		S_{n}\left({%
			\theta}\right)$}{TEXT} for all scoring rules}

The focused Bayesian update uses the term $S_{n}\left( {\theta }\right)
=\sum_{t=1}^{n}s\left( P_{{\theta }}^{(t-1)},y_{t}\right) $, thus
variational inference requires evaluation of 
\begin{equation*}
\nabla _{\theta }S_{n}\left( {\theta }\right) =\sum_{t=1}^{n}\nabla _{\theta
}s\left( P_{{\theta }}^{(t-1)},y_{t}\right) .
\end{equation*}%
For the mixture example we consider three alternative choices for $s\left(
P_{{\theta }}^{(t-1)},y_{t}\right) $, namely, the IS, the CLS and LS.
Here, we derive an expression for $\nabla _{\theta }s\left( P_{{\theta }%
}^{(t-1)},y_{t}\right) $ for each of these scores. As will be shown later,
the gradient $\nabla _{\theta }S\left( P_{{\theta }}^{(t-1)},y_{t}\right) $
for all the scores can be expressed solely in terms of $\nabla _{\theta
}p\left( y_{t}|\mathcal{F}_{t-1},{\theta }\right) $ and $\nabla _{\theta }P_{%
	{\theta }}^{(t-1)}$, thus we focus on the derivation of these two
expressions. Below, we denote $\epsilon _{t}=y_{t}-\mu $ and use the scalar $%
r$ to denote the number of elements in $\theta $. For ease of notation we
rewrite 
\begin{equation*}
p\left( y_{t}|\mathcal{F}_{t-1},{\theta }\right) =\frac{c_{2,t}}{c_{1,t}}%
\text{\ \ \ \ \ and \ \ \ \ \ }P_{{\theta }}^{(t-1)}=\frac{c_{3,t}}{c_{1,t}},
\end{equation*}%
where $c_{1,t,k}=\frac{\tau_{k}}{s_{k}}\phi _{1}\left( \frac{\epsilon
	_{t-1}-\mu _{k}}{s_{k}}\right) $, $c_{2,t,k}=\frac{1}{\sigma _{k}}\phi
_{1}\left( \frac{\epsilon _{t}-\beta _{k,0}-\beta _{k,1}\epsilon _{t-1}}{%
	\sigma _{k}}\right) $, $c_{3,t,k}=\Phi _{1}\left( \frac{\epsilon _{t}-\beta
	_{k,0}-\beta _{k,1}\epsilon _{t-1}}{\sigma _{k}}\right) $, $%
c_{1,t}=\sum_{k=1}^{K}c_{1,t,k}$, $c_{2,t}=\sum_{k=1}^{K}c_{1,t,k}c_{2,t,k}$
and $c_{3,t}=\sum_{k=1}^{K}c_{1,t,k}c_{3,t,k}$. The elements of the
gradients $\nabla _{\theta }p\left( y_{t}|\mathcal{F}_{t-1},{\theta }\right)
=\left[ \nabla _{\theta _{1}}p\left( y_{t}|\mathcal{F}_{t-1},{\theta }%
\right) ,\dots ,\nabla _{\theta _{r}}p\left( y_{t}|\mathcal{F}_{t-1},{\theta 
}\right) \right] ^{^{\prime }}$ and $\nabla _{\theta }P_{{\theta }}^{(t-1)}=%
\left[ \nabla _{\theta _{1}}P_{{\theta }}^{(t-1)},\dots ,\nabla _{\theta
	_{r}}P_{{\theta }}^{(t-1)}\right] ^{^{\prime }}$, can then be computed as 
\begin{equation}
\nabla _{\theta _{i}}p\left( y_{t}|\mathcal{F}_{t-1},{\theta }\right)
=\sum_{k=1}^{K}\left[ \tau_{k,t}\frac{\partial c_{2,t,k}}{\partial \theta
	_{i}}+\frac{c_{2,t,k}-p\left( y_{t}|\mathcal{F}_{t-1},{\theta }\right) }{%
	c_{1,t}}\frac{\partial c_{t,k,1}}{\partial \theta _{i}}\right]
\label{Eq:ypredictivegradient}
\end{equation}%
and 
\begin{equation}
\nabla _{\theta _{i}}P_{{\theta }}^{(t-1)}=\sum_{k=1}^{K}\left[ \tau_{k,t}%
\frac{\partial c_{3,t,k}}{\partial \theta _{i}}+\frac{c_{3,t,k}-P_{{\theta }%
	}^{(t-1)}}{c_{1,t}}\frac{\partial c_{t,k,1}}{\partial \theta _{i}}\right] .
\label{Eq:gradient}
\end{equation}%
Table \ref{tab:gradients} provides the expressions $\frac{\partial c_{1,t,k}%
}{\partial \theta _{i}}$, $\frac{\partial c_{2,t,k}}{\partial \theta _{i}}$
and $\frac{\partial c_{3,t,k}}{\partial \theta _{i}}$ for $\theta _{i}\in \{{%
	\beta }_{0},{\eta },{\kappa },\mu \}$. For $\theta _{i}\in {\psi }$, the
gradients can be evaluated as 
\begin{equation*}
\nabla _{\psi }p\left( y_{t}|\mathcal{F}_{t-1},{\theta }\right) =\left[
\nabla _{\tau}p\left( y_{t}|\mathcal{F}_{t-1},{\theta }\right) ^{\prime }%
\frac{\partial {\tau}}{\partial v}\frac{\partial {v}}{\partial \psi }\right]
^{\prime },
\end{equation*}%
and 
\begin{equation*}
\nabla _{\psi }P_{{\theta }}^{(t-1)}=\left[ {\nabla _{\tau}P_{{\theta }%
	}^{(t-1)}}^{\prime }\frac{\partial {\tau}}{\partial v}\frac{\partial {v}}{%
	\partial \psi }\right] ^{\prime },
\end{equation*}%
where $\frac{\partial {v}}{\partial \psi }$ is a diagonal matrix with
entries $\frac{\partial v_{k}}{\partial \psi _{k}}=\phi _{1}\left( \psi
_{k}\right) $, and $\frac{\partial {\tau}}{\partial v}$ is a lower
triangular matrix with diagonal elements $\frac{\partial \tau_{k}}{\partial
	v_{k}}=\prod_{j=1}^{k-1}(1-v_{j})$ and off-diagonal elements $\frac{\partial
	\tau_{k}}{\partial v_{s}}=-\frac{\tau_{k}}{(1-v_{s})}$, for $s<k$. The
expressions needed to evaluate $\nabla _{\tau}p\left( y_{t}|\mathcal{F}%
_{t-1},{\theta }\right) =\left[ \nabla _{\tau_{1}}p\left( y_{t}|\mathcal{F}%
_{t-1},{\theta }\right) ,\dots ,\nabla _{\tau_{K}}p\left( y_{t}|\mathcal{F}%
_{t-1},{\theta }\right) \right] ^{\prime }$ and $\nabla _{\tau}P_{{\theta }%
}^{(t-1)}=\left[ \nabla _{\tau_{1}}P_{{\theta }}^{(t-1)},\dots ,\nabla
_{\tau_{K}}P_{{\theta }}^{(t-1)}\right] ^{\prime }$ are also provided in
Table \ref{tab:gradients}.

With expressions \eqref{Eq:ypredictivegradient} and \eqref{Eq:gradient}, we
can now derive expression for $\nabla_\theta s\left(P_{{\theta}%
}^{(t-1)},y_{t}\right)$ for the three scores considered.

\subsubsection{Logarithmic score (LS)}

Denote the gradient of the LS in \eqref{ls} as 
\begin{equation*}
\nabla _{\theta }s^{\text{LS}}\left( P_{{\theta }}^{(t-1)},y_{t}\right)
=\left( \nabla _{\theta _{1}}s^{\text{LS}}\left( P_{{\theta }%
}^{(t-1)},y_{t}\right) ,\dots \right. \left. ,\nabla _{\theta _{r}}s^{\text{%
		LS}}\left( P_{{\theta }}^{(t-1)},y_{t}\right) \right) ^{\prime }.
\end{equation*}%
The element $\nabla _{\theta _{i}}s^{\text{LS}}\left( P_{{\theta }%
}^{(t-1)},y_{t}\right) $ of this gradient can be evaluated as 
\begin{equation*}
\nabla _{\theta _{i}}s^{\text{LS}}\left( P_{{\theta }}^{(t-1)},y_{t}\right) =%
\frac{1}{p\left( y_{t}|\mathcal{F}_{t-1},{\theta }\right) }\nabla _{\theta
	_{i}}p\left( y_{t}|\mathcal{F}_{t-1},{\theta }\right)
\end{equation*}%
where the derivative $\nabla _{\theta _{i}}p\left( y_{t}|\mathcal{F}_{t-1},{%
	\theta }\right) $ can be computed using (\ref{Eq:ypredictivegradient}).

\subsubsection{Censored logarithmic score (CLS)}

For some threshold value $y_{q}$, denote the upper tail support of
predictive distribution as $A=\{y_{t}:y_{t}>y_{q}\}$. The gradient for the
upper tail\textbf{\ }CLS in \eqref{cls} can be written as 
\begin{equation*}
\nabla _{\theta }s^{\text{CLS-A}}\left( P_{{\theta }}^{(t-1)},y_{t}\right)
=\left( \nabla _{\theta _{1}}s^{\text{CLS-A}}\left( P_{{\theta }%
}^{(t-1)},y_{t}\right) ,\dots ,\nabla _{\theta _{r}}s^{\text{CLS-A}}\left(
P_{{\theta }}^{(t-1)},y_{t}\right) \right) ^{\prime }.
\end{equation*}
We can compute the element $\nabla _{\theta _{i}}s^{\text{CLS-A}}\left( P_{{%
		\theta }}^{(t-1)},y_{t}\right) $ of the upper CLS as: 
\begin{equation*}
\nabla _{\theta _{i}}s^{\text{CLS-A}}\left( P_{{\theta }}^{(t-1)},y_{t}%
\right) =\nabla _{\theta _{i}}s^{\text{LS}}\left( P_{{\theta }%
}^{(t-1)},y_{t}\right) I(y_{t}\in A)+\frac{1}{P_{{\theta }}^{(t-1)}(y_{q})}%
\nabla _{\theta _{i}}P_{{\theta }}^{(t-1)}(y_{q})I(y_{t}\in A^{c}).
\end{equation*}%
The derivatives $\nabla _{\theta _{i}}P_{{\theta }}^{(t-1)}(y_{q})$ can be
computed using (\ref{Eq:gradient}). For the lower tail CLS, where $%
A=\{y_{t}:y_{t}<y_{q}\}$ we have that 
\begin{equation*}
\nabla _{\theta _{i}}s^{\text{CLS-A}}\left( P_{{\theta }}^{(t-1)},y_{t}%
\right) =\nabla _{\theta _{i}}s^{\text{LS}}\left( P_{{\theta }%
}^{(t-1)},y_{t}\right) I(y_{t}\in A)-\frac{1}{1-P_{{\theta }}^{(t-1)}(y_{q})}%
\nabla _{\theta _{i}}P_{{\theta }}^{(t-1)}(y_{q})I(y_{t}\in A^{c}).
\end{equation*}

\subsubsection{Interval score (IS)}

The gradient of the IS in \eqref{msis} can be written as 
\begin{equation*}
\nabla _{\theta }s^{\text{IS}}\left( P_{{\theta }}^{(t-1)},y_{t}\right)
=\left( \nabla _{\theta _{1}}s^{\text{IS}}\left( P_{{\theta }%
}^{(t-1)},y_{t}\right) \right. \left. ,\dots ,\nabla _{\theta _{r}}s^{\text{%
		IS}}\left( P_{{\theta }}^{(t-1)},y_{t}\right) \right) ^{\prime },
\end{equation*}
with elements $\nabla _{\theta _{i}}s^{\text{IS}}\left( P_{{\theta }%
}^{(t-1)},y_{t}\right) $ defined as: 
\begin{equation*}
\nabla _{\theta _{i}}s^{\text{IS}}\left( P_{{\theta }}^{(t-1)},y_{t}%
\right) =-\left[ 1-\frac{2}{1-q}I(y_{t}>u_{t})\right] \frac{\partial u_{t}}{%
	\partial \theta _{i}}-\left[ \frac{2}{1-q}I(y_{t}<l_{t})-1\right] \frac{%
	\partial l_{t}}{\partial \theta _{i}}
\end{equation*}%
As such, computation of $\nabla _{\theta _{i}}s^{\text{IS}}\left( P_{{%
		\theta }}^{(t-1)},y_{t}\right) $ entails evaluation of $\frac{\partial u_{t}%
}{\partial {\theta _{i}}}$ and $\frac{\partial l_{t}}{\partial {\theta _{i}}}
$. The derivative $\frac{\partial u_{t}}{\partial {\theta _{i}}}$ can be
evaluated by first noting that $\alpha _{u,t}=P_{{\theta }}^{(t-1)}(u_{t})$.
Then, using the triple product rule we know that 
\begin{equation*}
\frac{\partial u_{t}}{\partial \theta _{i}}=-\frac{\partial u_{t}}{\partial
	\alpha _{u,t}}\frac{\partial \alpha _{u,t}}{\partial \theta _{i}},
\end{equation*}%
where the first term can be computed as $\frac{\partial u_{t}}{\partial
	\alpha _{u,t}}=\frac{1}{p\left( u_{t}|\mathcal{F}_{t-1},{\theta }\right) }$.
The second term $\frac{\partial \alpha _{u,t}}{\partial \theta _{i}}=\nabla
_{\theta _{i}}P_{{\theta }}^{(t-1)}(u_{t})$ is evaluated using (\ref%
{Eq:gradient}). The derivative $\frac{\partial l_{t}}{\partial {\theta _{i}}}
$ is evaluated in the same fashion as described for $\frac{\partial u_{t}}{%
	\partial {\theta _{i}}}$.

\begin{table}[]
	\centering
	\resizebox{0.95\textwidth}{!}{
		\begin{tabular}{lll}
			\hline\hline
			&  &                                                                                                                                                                                                                                                                \\
			\multicolumn{3}{c}{Expressions for $\frac{\partial }{\partial\theta_i}c_{1,t,k}$}                                                                                                                                                                                                                           \\
			\multicolumn{3}{c}{\underline{\hspace{7cm}}}                                                                                                                                                                                                                                             \\
			&  &                                                                                                                                                                                                                                                                \\
			\multicolumn{3}{l}{$\frac{\partial c_{1,t,k}}{\partial \tau_{k}} = \frac{1}{s_k}\phi_1\left(\frac{\epsilon_{t-1}-\mu_k}{s_k}\right)$}                                                                                                                                                                                                                                                                                                                                                                                                   \\
			&  &                                                                                                                                                                                                                                                                \\
			\multicolumn{3}{l}{$\frac{\partial c_{1,t,k}}{\partial \beta_{0,k}} = -\frac{\tau_k}{s_k}\phi_1'\left(\frac{\epsilon_{t-1}-\mu_k}{s_k}\right)\frac{1}{(1-\beta_{k,1})s_k}$}                                                                                                                                                                                                                                                                                                                                                             \\
			&  &                                                                                                                                                                                                                                                                \\
			\multicolumn{3}{l}{$\frac{\partial c_{1,t,k}}{\partial \eta_{k}} =\left[ -\frac{\tau_k}{s_k}\phi_1'\left(\frac{\epsilon_{t-1}-\mu_k}{s_k}\right)\left(\frac{1}{s_k}\frac{\partial \mu_k}{\partial \beta_{1,k}}+\frac{1}{s_k^2}(\epsilon_{t-1}-\mu_k)\frac{\partial s_k}{\partial \beta_{1,k}}\right)-\phi_1\left(\frac{\epsilon_{t-1}-\mu_k}{s_k}\right)\frac{\tau_k}{s_k^2}\frac{\partial s_k}{\partial \beta_{1,k}}\right]\frac{\partial \beta_{1,k}}{\partial \eta_{k}}$}                                                               \\
			&  &                                                                                                                                                                                                                                                                \\
			\multicolumn{3}{l}{$\frac{\partial c_{1,t,k}}{\partial \kappa_{k}} = \left[\frac{\tau_k}{s_k}\phi_1'\left(\frac{\epsilon_{t-1}-\mu_k}{s_k}\right)\left(\frac{1}{s_k}\frac{\partial \mu_k}{\partial \sigma_k}+\frac{1}{s_k^2}(\epsilon_{t-1}-\mu_k)\frac{\partial s_k}{\partial \sigma_k}\right)-\phi_1\left(\frac{\epsilon_{t-1}-\mu_k}{s_k}\right)\frac{\tau_k}{s_k^2}\frac{\partial s_k}{\partial \sigma_k}\right]\frac{\partial \sigma_{k}}{\partial \kappa_{k}}$}                                                                      \\
			&  &                                                                                                                                                                                                                                                                \\
			\multicolumn{3}{l}{$\frac{\partial c_{1,t,k}}{\partial \mu} = -\frac{\tau_k}{s_k}\phi_1'\left(\frac{\epsilon_{t-1}-\mu_k}{s_k}\right)\frac{1}{s_k}$}                                                                                                                                                                                                                                                                                                                                                                            \\
			&  &                                                                                                                                                                                                                                                                \\ \hline
			&  &                                                                                                                                                                                                                                                                \\
			\multicolumn{1}{c}{Expressions for  $\frac{\partial }{\partial\theta_i}c_{2,t,k}$}                                                                                                                                                                               &  & \multicolumn{1}{c}{Expressions for  $\frac{\partial }{\partial\theta_i}c_{3,t,k}$}                                                                                                                                                                             \\
			\multicolumn{1}{c}{\underline{\hspace{7cm}}}                                                                                                                                                                                                                     &  & \multicolumn{1}{c}{\underline{\hspace{7cm}}}                                                                                                                                                                                                                   \\
			&  &                                                                                                                                                                                                                                                                \\
			$\frac{\partial c_{2,t,k}}{\partial \tau_k} = 0$                                                                                                                                                                                                                    &  & $\frac{\partial c_{3,t,k}}{\partial \tau_k} = 0$                                                                                                                                                                                                                  \\
			&  &                                                                                                                                                                                                                                                                \\
			$\frac{\partial c_{2,t,k}}{\partial \beta_{0,k}} = -\frac{1}{\sigma_k^2}\phi_1'\left(\frac{\epsilon_{t}-\beta_{0,k}-\beta_{1,k}\epsilon_{t-1}}{\sigma_k}\right)$                                                                                                 &  & $\frac{\partial c_{3,t,k}}{\partial \beta_{0,k}} = -\frac{1}{\sigma_k}\phi_1\left(\frac{\epsilon_{t}-\beta_{0,k}-\beta_{1,k}\epsilon_{t-1}}{\sigma_k}\right)$                                                                                                  \\
			&  &                                                                                                                                                                                                                                                                \\
			$\frac{\partial c_{2,t,k}}{\partial \eta_{k}} = -\frac{\epsilon_{t-1}}{\sigma_k^2}\phi_1'\left(\frac{\epsilon_{t}-\beta_{0,k}-\beta_{1,k}\epsilon_{t-1}}{\sigma_k}\right)\frac{\partial \beta_{1,k}}{\partial \eta_{k}} $                                        &  & $\frac{\partial c_{3,t,k}}{\partial \eta_{k}} = -\frac{\epsilon_{t-1}}{\sigma_k}\phi_1\left(\frac{\epsilon_{t}-\beta_{0,k}-\beta_{1,k}\epsilon_{t-1}}{\sigma_k}\right)\frac{\partial \beta_{1,k}}{\partial \eta_{k}} $                                         \\
			&  &                                                                                                                                                                                                                                                                \\
			$\frac{\partial c_{2,t,k}}{\partial \kappa_{k}} = -\frac{\epsilon_{t}-\beta_{0,k}-\beta_{1,k}\epsilon_{t-1}}{\sigma_k^3}\phi_1'\left(\frac{\epsilon_{t}-\beta_{0,k}-\beta_{1,k}\epsilon_{t-1}}{\sigma_k}\right)\frac{\partial \sigma_{k}}{\partial \kappa_{k}}-$ &  & $\frac{\partial c_{3,t,k}}{\partial \kappa_{k}} = -\frac{\epsilon_{t}-\beta_{0,k}-\beta_{1,k}\epsilon_{t-1}}{\sigma_k^2}\phi_1\left(\frac{\epsilon_{t}-\beta_{0,k}-\beta_{1,k}\epsilon_{t-1}}{\sigma_k}\right)\frac{\partial \sigma_{k}}{\partial \kappa_{k}}$ \\
			\hspace{1.5cm}$\frac{1}{\sigma_k^2}\phi_1\left(\frac{\epsilon_{t}-\beta_{0,k}-\beta_{1,k}\epsilon_{t-1}}{\sigma_k}\right)\frac{\partial \sigma_{k}}{\partial \kappa_{k}}$                                                                                                                                       &  &                                                                                                                                                                                                                                                                \\
			&  &                                                                                                                                                                                                                                                                \\
			$\frac{\partial c_{2,t,k}}{\partial \mu} = \frac{-1+\beta_{1,k}}{\sigma_k^2}\phi_1'\left(\frac{\epsilon_{t}-\beta_{0,k}-\beta_{1,k}\epsilon_{t-1}}{\sigma_k}\right)$                                                                              &  & $\frac{\partial c_{3,t,k}}{\partial \mu} = \frac{-1+\beta_{1,k}}{\sigma_k}\phi_1\left(\frac{\epsilon_{t}-\beta_{0,k}-\beta_{1,k}\epsilon_{t-1}}{\sigma_k}\right)$                                                                               \\
			&  &                                                                                                                                                                                                                                                                \\
			&  &                                                                                                                                                                                                                                                                \\ \hline\hline
	\end{tabular}	}
	\caption{{\protect\footnotesize Derivatives required to implement
		reparameterization trick for the mixture model. Other required expressions
		also include $\frac{\partial \protect\mu_k}{\partial \protect\beta_{1,k}} = 
		\frac{\protect\beta_{0,k}}{(1- \protect\beta_{1,k})^2}$, $\frac{\partial s_k%
		}{\partial \protect\beta_{1,k}} = \frac{\protect\beta_{1,k}\protect\sigma_k}{%
			(1-\protect\beta_{1,k}^2)^{3/2}} $, $\frac{\partial s_k}{\partial \protect%
			\sigma_k} = \frac{1}{(1-\protect\beta_{1,k}^2)^{1/2}}$, $\frac{\partial 
			\protect\sigma_k}{\partial \protect\kappa_k} = \exp\left(\protect\kappa%
		_k\right)$, $\frac{\partial \protect\beta _{1,k}}{\partial \protect\eta_k} =
		2\protect\phi_1(\protect\eta_k)$. The cross-component derivatives $\frac{%
			\partial c_{i,t,k}}{\partial \protect\tau_j}$, $\frac{\partial c_{i,t,k}}{%
			\partial \protect\beta_{0,j}}$, $\frac{\partial c_{i,t,k}}{\partial \protect%
			\eta_{j}}$, $\frac{\partial c_{i,t,k}}{\partial \protect\kappa_{j}}$ are all
		zero for $j\ne k$. Finally, in this table we denote $\protect\phi_1^{\prime
		}\left(x\right) = \frac{\partial \protect\phi_1\left(x\right)}{\partial x}$}%
	. }\label{tab:gradients}
\end{table}

\section{Bayesian Neural Network Predictive Class (Section \protect\ref{nn})}

\label{App:neuralnetwork}

\subsection{Priors}

The priors of the model parameters are set as $\omega _{k}\sim N(0,\text{Inf}%
)$ and $p(\sigma _{y}^{2})\propto \frac{1}{\sigma _{y}^{2}}$. The standard
deviation parameter $\sigma _{y}$ is transformed to the real line as $c=\log
\left( \sigma _{y}\right) $, so that the parameter vector is ${\theta }%
=\left( {\omega }^{\prime },c\right) ^{^{\prime }}$. Then, the prior density
can be written as $p({\theta })=p(c)\prod_{k=1}^{d}p(\omega _{k})$, where $%
p(\omega _{k})\propto 1$ and $p(c)\propto 1$. From this prior density we
have that $\nabla _{\theta }\log p\left( {\theta }\right) ={0}$.

\subsection{Derivation of \texorpdfstring{$\nabla_\theta
		S_{n}\left({%
			\theta}\right)$}{TEXT} for all scoring rules}

As in the previous appendix we can show that $\nabla_{\theta}S_{n}\left({%
	\theta}\right) = \sum_{t=1}^{n}\nabla_\theta s\left(P_{{\theta}%
}^{(t-1)},y_{t}\right).$ Thus, we must find an expression for $\nabla_\theta
S\left(P_{{\theta}}^{(t-1)},y_{t}\right)$ for each of the scores.

\subsubsection{Logarithmic score (LS)}

The gradient for the LS can be written as 
\begin{equation*}
\nabla _{\theta }s^{\text{LS}}\left( P_{{\theta }}^{(t-1)},y_{t}\right)
=\left( \nabla _{\omega }^{\text{LS}}s\left( P_{{\theta }}^{(t-1)},y_{t}%
\right) ^{\prime },\nabla _{c}s^{\text{LS}}\left( P_{{\theta }%
}^{(t-1)},y_{t}\right) \right) ^{\prime },
\end{equation*}%
with 
\begin{equation*}
\nabla _{\omega }s^{\text{LS}}\left( P_{{\theta }}^{(t-1)},y_{t}\right) =%
\frac{1}{\sigma _{y}^{2}}\left[ y_{t}-g\left( {z}_{t};{\omega }\right) %
\right] \nabla _{\omega }g\left( {z}_{t};{\omega }\right) \text{ and }\nabla
_{c}s^{\text{LS}}\left( P_{{\theta }}^{(t-1)},y_{t}\right) =-1+\frac{1}{%
	\sigma _{y}^{2}}\left[ y_{t}-g\left( {z}_{t};{\omega }\right) \right] ^{2}.
\end{equation*}%
The term $\nabla _{\omega }g\left( {z}_{t};{\omega }\right) $ can be
evaluated analytically through back propagation.

\subsubsection{Censored logarithmic score (CLS)}

For the upper tail CLS the gradient can be written as 
\begin{equation*}
\nabla _{\theta }s^{\text{CLS-A}}\left( P_{{\theta }}^{(t-1)},y_{t}\right)
=\nabla _{\theta }s^{\text{LS}}\left( P_{{\theta }}^{(t-1)},y_{t}\right)
I(y_{t}\in A)+\frac{1}{P_{{\theta }}^{(t-1)}(y_{q})}\nabla _{\theta }P_{{%
		\theta }}^{(t-1)}(y_{q})I(y_{t}\in A^{c}),
\end{equation*}%
where $\nabla _{\theta }P_{{\theta }}^{(t-1)}(y_{q})=\left( \nabla _{\omega
}P_{{\theta }}^{(t-1)}(y_{q})^{^{\prime }},\nabla _{c}P_{{\theta }%
}^{(t-1)}(y_{q})\right) ^{^{\prime }}$, $\nabla _{\omega }P_{{\theta }%
}^{(t-1)}(y_{q})=-\phi _{1}\left( y_{q};g\left( {z}_{t};{\omega }\right)
,\sigma _{y}^{2}\right) \nabla _{\omega }g\left( {z}_{t};{\omega }\right) $
and $\nabla _{c}P_{{\theta }}^{(t-1)}(y_{q})=-\phi _{1}\left( y_{q};g\left( {%
	z}_{t};{\omega }\right) ,\sigma _{y}^{2}\right) (y_{q}-g\left( {z}_{t};{%
	\omega }\right) )$. The gradient for the lower tail CLS is 
\begin{equation*}
\nabla _{\theta }s^{\text{CLS-A}}\left( P_{{\theta }}^{(t-1)},y_{t}\right)
=\nabla _{\theta }s^{\text{LS}}\left( P_{{\theta }}^{(t-1)},y_{t}\right)
I(y_{t}\in A)-\frac{1}{1-P_{{\theta }}^{(t-1)}(y_{q})}\nabla _{\theta }P_{{%
		\theta }}^{(t-1)}(y_{q})I(y_{t}\in A^{c}).
\end{equation*}

\subsubsection{Interval score (IS)}

The gradient of the IS can be written as 
\begin{equation*}
\nabla _{\theta }s^{\text{IS}}\left( P_{{\theta }}^{(t-1)},y_{t}\right)
=\left( \nabla _{\omega }s^{\text{IS}}\left( P_{{\theta }%
}^{(t-1)},y_{t}\right) \right. \left. ,\nabla _{c}s_{\text{IS}}\left( P_{{%
		\theta }}^{(t-1)},y_{t}\right) \right) ^{\prime },
\end{equation*}
with elements $\nabla _{\omega }s^{\text{IS}}\left( P_{{\theta }%
}^{(t-1)},y_{t}\right) $ and $\nabla _{c}s^{\text{IS}}\left( P_{{\theta }%
}^{(t-1)},y_{t}\right) $ defined as: 
\begin{align*}
\nabla _{\omega }s^{\text{IS}}\left( P_{{\theta }}^{(t-1)},y_{t}\right) &
=-\left[ 1-\frac{2}{1-q}I(y_{t}>u_{t})\right] \frac{\partial u_{t}}{\partial
	\omega }^{\prime }-\left[ \frac{2}{1-q}I(y_{t}<l_{t})-1\right] \frac{%
	\partial l_{t}}{\partial \omega }^{\prime } \\
\nabla _{c}s^{\text{IS}}\left( P_{{\theta }}^{(t-1)},y_{t}\right) & =-%
\left[ 1-\frac{2}{1-q}I(y_{t}>u_{t})\right] \frac{\partial u_{t}}{\partial c}%
-\left[ \frac{2}{1-q}I(y_{t}<l_{t})-1\right] \frac{\partial l_{t}}{\partial c%
}
\end{align*}%
The derivative $\frac{\partial u_{t}}{\partial {\omega }}$ can be evaluated
by first noting that $\alpha _{u,t}=P_{{\theta }}^{(t-1)}(u_{t})$. Then,
using the triple product rule we know that 
\begin{equation*}
\frac{\partial u_{t}}{\partial \omega }=-\frac{\partial u_{t}}{\partial
	\alpha _{u,t}}\frac{\partial \alpha _{u,t}}{\partial \omega },
\end{equation*}%
where the first term can be computed as $\frac{\partial u_{t}}{\partial
	\alpha _{u,t}}=\frac{1}{p\left( u_{t}|\mathcal{F}_{t-1},{\theta }\right) }$.
The second term $\frac{\partial \alpha _{u,t}}{\partial \omega }=\nabla
_{\omega }P_{{\theta }}^{(t-1)}(u_{t})$ is evaluated as in the subsection
above. The derivative $\frac{\partial l_{t}}{\partial {\omega }}$ is
evaluated in the same fashion as described for $\frac{\partial u_{t}}{%
	\partial {\omega }}$. The corresponding derivatives for parameter $c$ can
also be computed using similar steps.

\end{document}